\documentclass[twocolumn]{autart}

\usepackage{graphics}
\usepackage{cite}
\usepackage{amsmath}
\usepackage{amssymb}
\usepackage{amsxtra}
\usepackage{arydshln}
\usepackage{amsfonts}
\usepackage[mathscr]{eucal}
\usepackage{mathrsfs}
\usepackage{psfrag}
\usepackage{caption}
\usepackage{subcaption}
\usepackage{url}
\usepackage[pdftex]{graphicx}
\usepackage{multirow}
\usepackage{url}
\usepackage{color}
\usepackage{interval}
\usepackage{hyperref}
\hypersetup{colorlinks = true,
            linkcolor = blue,
            urlcolor  = blue,
            citecolor = black,
            anchorcolor = blue}

\def\Real{\mathbb{R}}

\newcommand{\Xbar}{\bar{\mathcal{X}}}
\newcommand{\xbar}{\bar{{x}}}
\newcommand{\Ltwoe}{\boldsymbol{\rm L}_{2e}}
\newcommand{\Ltwop}{\boldsymbol{\rm L}_{2+}}
\newcommand{\graphe}[1]{\mathcal{G}_\epsilon\left(#1\right)}
\newcommand{\graphei}[1]{\mathcal{G}_\epsilon^\prime\left(#1\right)}

\newcommand{\TwoOne}[2]
{\begin{bmatrix}
{#1} \\
{#2}
\end{bmatrix}
}
\newcommand{\OneTwo}[2]
{\begin{bmatrix} {#1} & {#2}
\end{bmatrix}
}
\newcommand{\TwoTwo}[4]
{\begin{bmatrix}
{#1} & {#2} \\
{#3} & {#4}
\end{bmatrix}
}
\newcommand{\STwoOne}[2]
{\left[\begin{smallmatrix}
{#1} \\
{#2}
\end{smallmatrix}\right]
}
\newcommand{\SThreeOne}[3]
{\left[\begin{smallmatrix}
{#1} \\
{#2} \\
{#3}
\end{smallmatrix}\right]
}
\newcommand{\STwoTwo}[4]
{\left[\begin{smallmatrix}
{#1} & {#2} \\
{#3} & {#4}
\end{smallmatrix}\right]
}
\newcommand{\SOneTwo}[2]
{\left[\begin{smallmatrix}
{#1} & {#2}
\end{smallmatrix}\right]
}

\newcommand{\diag}{\mathrm{diag}}

\newcommand{\Ltwo}{\boldsymbol{\rm L}_{2}}

\newcommand{\norm}[1]{\left\lVert#1\right\rVert}

\intervalconfig{soft open fences}

\edef\endfrontmatter{
  \unexpanded\expandafter{\endfrontmatter} 
  \noexpand\endNoHyper}

\begin{document}
\begin{frontmatter}
\title{Feedback Stability Analysis via Dissipativity with  \\ Dynamic Supply Rates \thanksref{footnoteinfo}}
\thanks[footnoteinfo]{This paper was not presented at any IFAC
meeting. This work was supported in part by   the Research Grants
Council of Hong Kong, China under the General Research
Fund No.~16203922,  the European Research Council under the
Advanced ERC Grant Agreement SpikyControl n.101054323, and the Engineering and Physical Sciences Research Council (EPSRC) [grant number EP/R008876/1]. All research data supporting this publication are directly available within this publication. For the purpose of open access, the authors have applied a Creative Commons Attribution (CC BY) licence to any Author Accepted Manuscript version arising.}
\author[SZ]{Sei~Zhen~Khong}\ead{szkhong@mail.nsysu.edu.tw},  
\author[KU,CC]{Chao~Chen}\ead{cchenap@connect.ust.hk}, 
\author[AL]{Alexander~Lanzon}\ead{Alexander.Lanzon@manchester.ac.uk}
\address[SZ]{Department of Electrical Engineering, National Sun Yat-sen University, Kaohsiung 80421, Taiwan} 
\address[KU]{Department of Electrical Engineering (ESAT), KU Leuven, Kasteelpark Arenberg 10, 3001 Leuven, Belgium} 
\address[CC]{Department of Electronic and Computer Engineering, The Hong Kong University of Science and Technology, \\Clear Water Bay, Kowloon, Hong Kong, China} 
\address[AL]{Department of Electrical and Electronic Engineering, School of Engineering, University of Manchester, \\ Manchester M13 9PL, UK}           
\thanks{Corresponding author: C. Chen}  

\begin{abstract}
We propose a general notion of dissipativity with dynamic supply rates for nonlinear systems. This extends classical dissipativity with static supply rates and dynamic supply rates of miscellaneous quadratic forms. The main results of this paper concern Lyapunov and asymptotic stability analysis for nonlinear feedback dissipative systems that are characterised by dissipation inequalities with respect to compatible dynamic supply rates but involving possibly different and independent auxiliary systems. Importantly, dissipativity conditions guaranteeing stability of the state of the feedback systems, without concerns on the stability of the  state of the auxiliary systems, are provided. The key results also specialise to a simple coupling test for the interconnection of two nonlinear systems described by dynamic $(\Psi, \Pi, \Upsilon, \Omega)$-dissipativity, and are shown to recover several existing results in the literature, including small-gain, passivity indices, static $(Q, S, R)$-dissipativity, dissipativity with terminal costs, etc. Comparison with the input-output approach to feedback stability analysis based on integral quadratic constraints is also made.
\end{abstract}
\end{frontmatter}

\section{Introduction}
The notion of dissipativity of dynamical systems was first introduced by Jan C. Willems in~\cite{Wil72a, Wil72b}. The seminal work has profoundly
influenced   research  in the systems and control community, so much so that the \textit{IEEE Control Systems Magazine} recently published a
special two-part issue~\cite{Dis22a, Dis22b} commemorating the 50$^\text{th}$ anniversary of the papers~\cite{Wil72a, Wil72b}. Dissipativity theory abstracts
the notion of energy and its dissipation in dynamical systems, and may be viewed as a generalisation of Lyapunov theory for autonomous systems to open
systems with input and outputs~\cite{Wil07history}. Importantly, the construction of storage functions for linear systems with quadratic supply rates
led to the emergence of linear matrix inequalities~\cite{BGFB94} in the field of control. The literature on dissipativity is vast. It
incorporates~\cite{HilMoy76, MoyHil78, HilMoy80, Hill22} on stability theory and basic properties, \cite{Wil73, Schaft21} on cyclo-dissipativity, \cite{Wil07} on
behavioural systems, \cite{StaSep07} on synchronisation of nonlinear oscillators, \cite{Griggs09} on mixed small-gain/passive systems, and~\cite{Sepulchre:22,Liu:23, Verhoek:23} on incremental dissipativity, to name a few. Several books have also been written on the
topic~\cite{BaoLee07, BLME07, Sch17}. The importance of dissipativity to the field of systems and control is thus self-evident from the large existing
literature on it. Not only is dissipativity a notion that is physically motivated by energy dissipation, it is also interrelated with a broad range of systems
tools including Riccati equations~\cite{Wil72b, ZDG96}, the Kalman--Yakubovich--Popov lemma~\cite{Ran96, HilMoy80}, the Iwasaki--Hara lemma~\cite{IwaHar05} and linear matrix inequalities. When applied
to robust closed-loop stability analysis, dissipativity theory unifies the small-gain and passivity results~\cite{Vid02}.

Traditionally, dissipativity has been defined in terms of static and time-invariant supply rates and dynamics are confined to the corresponding
storage functions of the states of the  dynamical input-state-output systems in question.  When it comes to robust closed-loop (asymptotic/exponential) stability
analysis based on dissipativity, conservatism may be reduced with the aid of stable and stably invertible dynamical multipliers~\cite[Sec.~3.5.1]{GreLim95}. This method suffers
from arguably serious drawbacks in that the need for the multipliers to be stable and stably invertible substantially restricts their usefulness. On
the contrary, the input-output approach to robust feedback (input-output finite-gain) stability analysis~\cite{Zam66, DesVid75} has been shown to
naturally accommodate a wider class of multipliers in a direct fashion, for example, via the notion of integral quadratic constraints (IQCs)~\cite{MegRan97}. Under well-posedness assumptions, graph separation is necessary and sufficient for input-output closed-loop stability~\cite{DGS93, Tee96, TGPS11, Hilborne22}. The theory of IQCs~\cite{MegRan97, RanMeg97, CJKh13, CarSei19, Kho22} provides a powerful and unifying framework for establishing graph separation, and thus input-output closed-loop stability,~e.g.~\cite{Khong:18, Zhao:22, KhongLanzon24}. The type of multipliers that may be accommodated in this theory is extensive. It includes Zames-Falb~\cite{ZamFal68}, small-gain, passivity, $(Q, S, R)$-dissipativity, and circle criterion, to mention just a few. In the linear time-invariant (LTI) setting, it is also known that IQCs are nonconservative for robust stability analysis~\cite{IwaHar98, KhoKao21, KhoKao22, RMCQK22}.

Motivated in part by the prowess and utility of multipliers, the notion of dissipativity with dynamic supply rates has been considered for robust
stability analysis in various contexts \cite{XHK05, Sei15, AMP16, Sch21, WilTre98, WilTre02, Forni18, Angeli06, Schaft13, Ghallab22, Lanzon23, Bhowmick24}. In
\cite{WilTre98}, dynamic supply rates of quadratic differential forms are considered and physically motivated by numerous
examples. Dynamic supply rates of quadratic forms based on either affine nonlinear or LTI auxiliary systems are investigated in \cite{XHK05, Sei15,
  AMP16, Sch21}. Notably, dynamic supply rates for differential dissipativity \cite{Forni:13, Forni18, Verhoek:23}, differential passivity \cite{Schaft13},
counterclockwise dynamics\cite{Angeli06}, negative imaginariness   \cite{Ghallab22,Lanzon23,Bhowmick24} and system phase\cite{Chen:20} have also been examined in the literature.

In this paper, we propose a dissipativity notion involving dynamic supply rates of general form for nonlinear dynamical systems. The proposed
notion generalises static supply rates and  the  dynamic supply rates of  quadratic (and differential) forms considered in
\cite{XHK05, Sei15, AMP16, Sch21, WilTre98, WilTre02}, and may be used to capture the class of input-output negative imaginary systems
in~\cite{Lanzon23, Bhowmick24}. It relies on the dynamics of a possibly nonlinear auxiliary system that may be independent of the dynamics in the supply rate, marking a
remarkable departure from the literature. The dynamics in the supply rate may be seen as counterparts to the multipliers used in the definitions of
IQCs, whereas the dynamics of the auxiliary system facilitate the verification of the dissipativity of the system with respect to the supply rate in
question.

Lyapunov stability  and asymptotic  stability of a feedback interconnection consisting of two nonlinear dissipative systems sharing the same dynamic supply rate and utilising possibly different auxiliary systems are established in this paper. The use of dynamic supply rates
has the   advantage of reducing conservatism in feedback stability analysis similar to using multipliers in  input-output
stability analysis. An interesting specialisation of the key results of this paper is a simple coupling test to check the feedback stability of  two nonlinear systems that are described by  dynamic $(\Psi, \Pi, \Upsilon, \Omega)$-dissipation inequalities. We also specialise our main results to several existing results in the literature \cite{Sch17, Sch21, XHK05} on static and dynamic
supply rates of miscellaneous quadratic forms. Last, but not least, it is noteworthy that in many aspects, our main results are distinct from the IQC result with incrementally
bounded multipliers for input-output closed-loop stability analysis~\cite{Kho22}, and  their differences and relation are also discussed in detail.

The remainder of this paper is organised as follows. In Section~\ref{sec: dynamic_diss}, we propose a novel general notion of dissipativity with dynamic supply
rates and provide illustrating examples of systems that it can capture. The main results on feedback Lyapunov and asymptotic stability via the newly proposed general dissipativity notion are given 
in Section~\ref{sec: FB}. Section~\ref{sec: examples} shows that  several existing results in the literature can be stated as  corollaries of the main results in this paper. Section~\ref{sec:
  relation_IQCs} provides the differences and relation between the main results and IQC based feedback input-output stability. Section~\ref{sec: simu} provides a numerical example to demonstrate the utility of the main results and Section~\ref{sec: conclusion}
summarises this paper.

\textit{Notation}: Let $\mathbb{N}$, $\mathbb{R}$, $\Real_+$, $\mathbb{R}^n$ and $\mathbb{R}^{p\times m}$ denote the sets of natural numbers excluding $0$, real numbers, nonnegative real numbers,
$n$-dimensional real vectors, and $p\times m$ real matrices, respectively.  Given a matrix $M$, its transpose is denoted by $M^\top$.  An identity matrix of compatible size is denoted by $I$. Let $\norm{x}= \sqrt{x^\top x}$ for $x \in \Real^n$. A function $\alpha : \Real^n \to \Real$ is said to be positive definite if
$\alpha(0)=0$ and $\alpha(r)>0$ for all $0 \neq r \in \Real^n$. A function $\alpha : \Real_+ \to \Real_+$ is said to belong to class
$\mathcal{K}$ if it is continuous, strictly increasing, and $\alpha(0) = 0$. It is said to belong to class $\mathcal{K}_\infty$ if it belongs to class
$\mathcal{K}$ and $\displaystyle\lim_{r \to \infty} \alpha(r) = \infty$.  The extended space of $\Real$-valued Lebesgue absolutely integrable
functions is defined as
 \[ 
  \boldsymbol{\rm L}_{1e}   = \Big\{v : \Real_+ \to \Real \ \Big|\   \int_{0}^T \|v(t)\| \, dt< \infty \; \forall T \in [0, \infty) \Big\}. 
 \]
 An operator $\Psi$ maps an input $u$ in some signal space to an output $y$ in another space  via $y=\Psi(u)$. An operator can capture any static, dynamic, linear or nonlinear system. Define the truncation operator $(P_Tu)(t) = u(t)$ for
$t \leq T$ and $(P_Tu)(t) = 0$ for $t > T$. An operator $\Psi$ is said to be \emph{causal} if $P_T \Psi P_T = P_T\Psi$ for all $T \geq 0$. We denote an LTI system 
 \begin{align*}
   \dot{x}(t) & = Ax(t) + Bu(t) \\
   y(t) &= Cx(t) + Du(t)
 \end{align*}
 by its realisation $(A, B, C, D)$, where $A, B, C$ and $D$ are  real matrices with compatible dimensions. 
  
 \section{Dissipativity with Dynamic Supply Rates}\label{sec: dynamic_diss}
 
All dynamics considered in this paper are time-invariant. Consider a nonlinear input-state-output system
\begin{equation}\label{eq: OLSystem}
  \Sigma:\;
  \begin{aligned}
    \dot{x}(t) &= f(x(t), u(t)), \; x(0) = x_0, \; x(t) \in \mathcal{X}, \; u(t) \in \mathcal{U} \\
    y(t) &= h(x(t), u(t)), \; y(t) \in \mathcal{Y},
  \end{aligned}
\end{equation}
with $\mathcal{X}=\Real^n$, $\mathcal{U}=\Real^m$, $\mathcal{Y} = \Real^p$, locally Lipschitz  $f:\mathcal{X}\times \mathcal{U}\to \mathcal{X}$ and continuous $h:\mathcal{X}\times \mathcal{U}\to \mathcal{Y}$.    An input $u$ to $\Sigma$ in \eqref{eq:
  OLSystem} is called \emph{admissible} if there exists a unique solution $x(t)$ on $t\in [0, \infty)$ for every initial condition $x(0)$. The set of
all admissible inputs to $\Sigma$ is denoted by $\mathscr{U}$ and the set of outputs of $\Sigma$ over $\mathscr{U}$ is denoted as $\mathscr{Y}$.  Based
on the admissible sets $\mathscr{U}$ and $\mathscr{Y}$, we propose the following definition of a dynamic supply rate:
 
\begin{defn}[Dynamic supply rate]
  A time function $\xi$  is called a \emph{dynamic
  supply rate} for $\Sigma$ if it is the output  of a causal time-invariant dynamic  operator 
\begin{equation}\label{eq: srate_map}
  \Xi: \mathscr{U} \times \mathscr{Y} \times {\bar{\mathcal{X}}}  \to \boldsymbol{\rm L}_{1e}.
\end{equation}
In other words,  $\xi(t) = \Xi(u, y, \bar{x})(t)$, where $u\in \mathscr{U}$, $y \in \mathscr{Y}$, and $\bar{x}\in \mathcal{\bar{X}}=\mathbb{R}^{n_{\bar{x}}}$.   
\end{defn}

 Note that the operator $\Xi$ in \eqref{eq: srate_map} can capture any causal system (whether it is static or dynamic, linear or nonlinear). A characterisation of $\Xi$ via, for example, state-space equations, is not necessary. The argument $\bar{x} \in \mathcal{\bar{X}}$ in \eqref{eq: srate_map} is only used for specifying initial conditions and is significant for the following reason. In the case where $\Xi$ has a state-space representation, its initial condition may be taken to be an arbitrary function of $\bar{x}\in \bar{\mathcal{X}}$. In the case where $\Xi$ has no state-space representation or the initial condition of its state-space representation is fixed (e.g.~at $0$), the dependence on $\bar{x}$ is inconsequential.

Throughout this paper, the dynamic supply rate $\xi$ and its associated  operator $\Xi$ may be used interchangeably without ambiguity since they represent the same object. A supply rate $\xi(t)  = \Xi(u,y,\xbar)(t)$ is \emph{static} when there exists $\tilde{\Xi} : \Real^{m}\times \Real^p \to \Real$ such that $\xi(t)  = \Xi(u,y,\xbar)(t) = \tilde{\Xi}(u(t), y(t))$ for all $t\geq 0$, $u\in \mathscr{U}$, $y \in \mathscr{Y}$ and $\xbar\in\Xbar$, i.e. the supply rate is independent of $\xbar\in\Xbar$ and the dependency on $u\in \mathscr{U}$ and $y \in \mathscr{Y}$ is that of a static function.

An example of a dynamic supply rate $\xi$ of the {quadratic} form is
\begin{align} \label{eq: quadratic_supply}
  \xi(t) = \Xi(u, y, \bar{x})(t) = \left(\Psi \TwoOne{u}{y}\right)(t) ^\top \left(\Pi \TwoOne{u}{y}\right)(t),
\end{align}
where $\Psi$ and $\Pi$ are two causal,  nonlinear, dynamic operators whose initial conditions may depend on $\bar{x}$. This is a generalisation of the supply rates considered in~\cite{XHK05, Sei15, AMP16, Sch21}, where $\Psi$ and $\Pi$ are taken to be either
affine nonlinear or linear time-invariant operators and $\bar{\mathcal{X}}=\emptyset$. Another example of a dynamic supply rate is the quadratic differential form considered
in~\cite{WilTre98, WilTre02}:   
\[
  \xi(t) = \Xi(u, y, {\xbar})(t)= \sum_{k, \ell} \left(\frac{d^k \STwoOne{u}{y}}{ dt^k} (t)\right) ^\top P_{k\ell} \left(\frac{d^\ell
      \STwoOne{u}{y}}{ dt^\ell} (t)\right),
\]
where $P_{k\ell} \in \Real^{(m+p) \times (m+p)}$. 

For $\Sigma$ in \eqref{eq: OLSystem},  we associate with it an \emph{auxiliary system}\footnote[1]{One may adopt the form $\dot{z}(t) = \hat{g}(z(t), y(t), u(t))$ and $\phi(t) = \hat{h}_\Phi(z(t), y(t), u(t))$, which is a special case of \eqref{eq: aux} by noting $y(t)=h(x(t), u(t))$.} 
\begin{equation} \label{eq: aux}
\Phi:\; 
\begin{aligned}
  \dot{z}(t) &= g(z(t), x(t), u(t)), \; z(0)=z_0, z(t) \in \mathcal{Z} \\  
    \phi(t) & =  h_\Phi(z(t), x(t), u(t)),\; \phi(t) \in \mathcal{O},
  \end{aligned}
\end{equation}
  with $\mathcal{Z}=\Real^{n_z}$ and $\mathcal{O}=\Real^{p_{\phi}}$, locally Lipschitz $g: \mathcal{Z}\times\mathcal{X}\times\mathcal{U}\to \mathcal{Z}$ and continuous $h_\Phi: \mathcal{Z}\times\mathcal{X}\times\mathcal{U}\to \mathcal{O}$. It is assumed throughout that there exists a unique solution $z(t)$ on $t\in [0, \infty)$ to
  \eqref{eq: aux} for all $u \in \mathscr{U}$.    Equipped with the dynamic supply rates and auxiliary systems, we are ready to generalise the
  classical notion of dissipativity.
 
\begin{defn}[Dynamic dissipativity]\label{def: dynamic_dissi} 
  Let $\Xi: \mathscr{U} \times \mathscr{Y}  \times \bar{\mathcal{X}}  \to \boldsymbol{\rm L}_{1e}$ be causal. 
  $\Sigma$ in \eqref{eq: OLSystem} is said to be  \emph{$\Xi$-dissipative on $(\mathcal{X}, \mathscr{U})$} if there exist an auxiliary system  \eqref{eq: aux} and a \emph{storage function} $S : \mathcal{X} \times \mathcal{Z} \to \Real$ such that the following \emph{dissipation inequality}
  \begin{equation}\label{eq: dissipation}
    \begin{aligned}
    S(x(T), z(T)) \leq S(x(0), z(0))  + \int_{0}^{T} \xi(t) \, dt
    \end{aligned}
  \end{equation}
   holds for all $T > 0$, $u\in \mathscr{U}$, $x(0) \in \mathcal{X}$, and $\bar{x} \in \mathcal{\bar{X}}$, where $\xi(t)= \Xi(u, y, \bar{x})(t)$ and $x, z, y$ satisfy \eqref{eq: OLSystem} and \eqref{eq: aux}. Furthermore, $\Sigma$ is said to be \emph{$\Xi'$-dissipative} if \eqref{eq: dissipation} holds for all $T > 0$, $u\in \mathscr{U}$, and $x(0) \in \mathcal{X}$, where $\xi(t)= \Xi(u, y, x(0))(t)$ and $x, z, y$ satisfy \eqref{eq: OLSystem} and \eqref{eq: aux}.
\end{defn}

Note from the definition above that   $\Xi'$-dissipativity is an \emph{easier} property to satisfy than $\Xi$-dissipativity, since the former is required to hold only for $\bar{x} = x(0)$ in the supply rate while the latter needs to hold for all $\bar{x} \in \mathcal{\bar{X}}$, independently of $x(0) \in \mathcal{X}$. The purpose of defining both $\Xi$-dissipativity and $\Xi'$-dissipativity will be made clear when feedback stability is examined in Section~\ref{sec: FB}. 
When the chosen supply rate $\Xi(u,y,\xbar)$ is independent of $\xbar$, the two notions $\Xi$-dissipativity and $\Xi'$-dissipativity are identical.

When $\Sigma$ is static, $\mathcal{X} = \emptyset$ and therefore only $\Xi$-dissipativity is sensible in Definition~\ref{def: dynamic_dissi} with $S(z)$ replacing $S(x,z)$ in \eqref{eq: dissipation}, i.e.~a static system $\Sigma$ is $\Xi$-dissipative if there exist an auxiliary system~\eqref{eq: aux} and $S : \mathcal{Z} \to \Real$ such that
\begin{align}\label{eq: static_dissi_ineq}
  S(z(T)) \leq S(z(0)) + \int_{0}^{T} \xi(t) \, dt
\end{align}
for all $T>0$, $u\in \mathscr{U}$ and $\bar{x} \in \mathcal{\bar{X}}$, where $\xi(t)= \Xi(u, y, \bar{x})(t)$.

\begin{figure}[htb]
  \centering
  \setlength{\unitlength}{1.1mm}
  \begin{picture}(50,43)
  \thicklines
  \put(7,10){\vector(1,0){10.5}}
  \put(10,39){\vector(1,0){30}}
  \put(2,25){\vector(1,0){13}}
  \put(15,20){\framebox(15,10){$\Sigma$}}
  \put(30,25){\vector(1,0){15}}  
  \put(22.5,20){\vector(0,-1){5}}
   \put(17.5,5){\framebox(10,10){$\Phi$}}
   \put(29, 10){\makebox(5,5){$\phi$}} 
   \put(27.5,10){\vector(1,0){7}}
   \put(40,32.5){\framebox(10,10){$\Xi$}}
   \put(50,37.5){\vector(1,0){7}}
   \put(51.5, 37.5){\makebox(5,5){$\xi$}} 
   \put(17.5,0){\makebox(5,5){$z$}} 
   \put(22.5,5){\vector(0,-1){5}}
   \put(36,35){\vector(1,0){4}}
  \put(10,25){\line(0,1){14}}
  \put(36,25){\line(0,1){10}}
  \put(7,10){\line(0,1){15}}
   \put(24,31){\makebox(5,5){$x(0)$}} 
     \put(39,37){\vector(1,0){1}}
    \put(38.5,37){\line(1,0){1}}
    \put(36.5,37){\line(1,0){1}}
    \put(34.5,37){\line(1,0){1}}
     \put(32.5,37){\line(1,0){1}}
    \put(30.5,37){\line(1,0){1}}
    \put(28.5,37){\line(1,0){1}}
     \put(26.5,37){\line(1,0){1}}
    \put(24.5,37){\line(1,0){1}}
      \put(22.5,37){\line(1,0){1}}
       \put(22.5,37){\line(0,-1){1}}
    \put(22.5,35){\line(0,-1){1}}
    \put(22.5,33){\line(0,-1){1}}
    \put(22.5,31){\line(0,-1){1}}
  \put(17.5,15){\makebox(5,5){$x$}} 
  \put(31.5,25){\makebox(5,5){$y$}}
  \put(3,25){\makebox(5,5){$u$}} 
  \end{picture}\caption{A physical system $\Sigma$ with state variable $x$, input $u$ and output $y$.  It is  associated with a dynamic supply rate $\xi=\Xi(u, y, \xbar)$ and  an auxiliary system $\Phi$ with state variable $z$ and output $\phi$. The dotted line signifies that the initial condition of $\Xi$ may be taken to be a function of $\xbar=x(0)$.}\label{fig: aux}
  \end{figure}
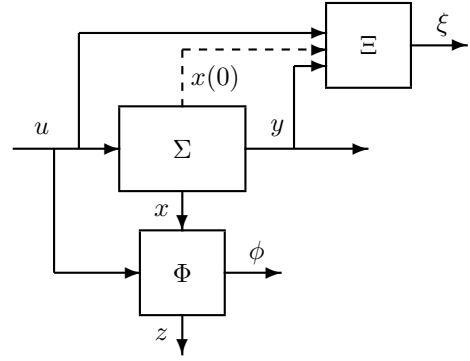

  The auxiliary system $\Phi$ in Definition~\ref{def: dynamic_dissi} may be related to the dynamics in the operator $\Xi$. It may also be  empty with $\mathcal{Z} = \emptyset$,  in which case we have $S : \mathcal{X} \to \Real$ and an obvious simplification in Definition~\ref{def: dynamic_dissi}.   In general, the auxiliary system $\Phi$ and dynamic
  operator $\Xi$ may be independent of each other. See Fig.~\ref{fig: aux} for an illustration. One notably motivating
    example  involves the class of input-output negative imaginary (IONI) systems \cite{Lanzon23}, which is
    elaborated in Example~\ref{exmp: IONI}.  The purpose of introducing an auxiliary
    system \eqref{eq: aux}  is to facilitate the verification of the dissipation inequality \eqref{eq: dissipation} with a dynamic supply rate.  This idea enjoys certain similarities with that of using prolonged systems \cite{Crouch:87} in differential dissipativity variational analysis in \cite[Def.~3]{Forni:13} and \cite{Verhoek:23}, but is also fundamentally different  since the present paper is concerned with dissipativity as opposed to its differential or incremental form \cite[Sec.~2]{Sepulchre:22}.

If $S$ is continuously differentiable, we say that it is a $C^1$ storage function. To illustrate the verification of dissipativity with dynamic supply rates, we note the following lemma when $S$ is a $C^1$ storage function.

\begin{lem}\label{prop: differential_dissi}
 Let $\Sigma$ be given by \eqref{eq: OLSystem} and the auxiliary system $\Phi$ be given by \eqref{eq: aux}. Then $\Sigma$ is {$\Xi$-dissipative} with a $C^1$ storage function $S$ if 
 \begin{multline}\label{eq: differential_dissi}
 \hspace{-3mm}{\left(\left[\frac{\partial}{\partial x} S(x, z)\right]^\top f(x, u)\right)(t)}  \\
 {+  \left(\left[\frac{\partial}{\partial z} S(x, z)\right]^\top g(z, x, u)\right)(t)  \leq \xi(t)}
 \end{multline}
 for all $t\geq 0$, $u\in \mathscr{U}$, $x(0) \in \mathcal{X}$ and $\bar{x} \in \mathcal{\bar{X}}$, where $\xi(t)=\Xi(u, y, \bar{x})(t)$ and $x$, $z$,   $y$ satisfying \eqref{eq: OLSystem} and \eqref{eq: aux}. Similarly, $\Sigma$ is $\Xi'$-dissipative if \eqref{eq: differential_dissi} holds with $\xi(t)= \Xi(u, y, x(0))(t)$.
\end{lem}
\pf
The proof follows from Definition~\ref{def: dynamic_dissi} by taking integrals on both sides of \eqref{eq: differential_dissi}. \hfill $\square$
\endpf
 
Assuming that $\Xi$  has a  state-space representation which shares the same state equation with that of the auxiliary system  $\Phi$ in \eqref{eq: aux}, then \eqref{eq: differential_dissi} may be verified via an algebraic inequality. Specifically, let
 $\Phi$ be LTI with realisation $(A_\Phi, B_\Phi, C_\Phi, D_\Phi)$ and the supply rate be  $\xi(t)=\Xi(u,
  y, \bar{x})(t)=\left(\Phi\STwoOne{u}{y}\right)(t)^\top P \left(\Phi\STwoOne{u}{y}\right)(t)$, where $P\in \Real^{(m+p) \times (m+p)}$ and the initial condition of $\Phi$ may depend on $\bar{x}\in\bar{\mathcal{X}}$, then \eqref{eq:
    differential_dissi} holds if
    \begin{multline*}
      \hspace{-5mm}\left[\frac{\partial}{\partial x} S(x, z)\right]^\top f(x, u)  +   \left[\frac{\partial}{\partial z} S(x, z)\right]^\top \left(A_\Phi z + B_\Phi \TwoOne{u}{h(x,u)}\right)\\
           \begin{aligned}
            \leq \left( C_\Phi z + D_\Phi \TwoOne{u}{h(x, u)} \right)^\top P \left( C_\Phi z + D_\Phi \TwoOne{u}{h(x, u)} \right)
           \end{aligned}
      \end{multline*}
      for all $x\in \mathcal{X}$, $z\in \mathcal{Z}$ and $u\in\mathcal{U}$.
   In the case where $\Phi$ is stable, this coincides with that considered in~\cite[Sec.~8.1]{AMP16}. More generally, let the state-equation of the auxiliary system $\Phi$ be $\dot{z}(t) = \hat{g}(z(t), y(t), u(t))$, where $y(t)=h(x(t), u(t))$. Let the supply
  rate $\xi(t)$ be of the general quadratic form \eqref{eq: quadratic_supply}, where the outputs of causal operators
  $\Psi = \STwoOne{u}{y} \mapsto {\phi_1}$ and $\Pi = \STwoOne{u}{y} \mapsto {\phi_2}$ are described by $\phi_1(t) = \hat{h}_{\Phi_1}(z(t), y(t), u(t))$ and
  $\phi_2(t) = \hat{h}_{\Phi_2}(z(t), y(t), u(t))$, respectively, and they share the same state-equation involving $z$ as in $\Phi$, i.e.~the output of $\Phi$ is $\STwoOne{\phi_1}{\phi_2}$. Then, \eqref{eq: differential_dissi} holds if
\begin{multline*}
  \left[\frac{\partial}{\partial x} S(x, z)\right]^\top f(x, u)  + \left[\frac{\partial}{\partial z} S(x, z)\right]^\top  \hat{g}(z, h(x, u), u) \\
  \begin{aligned}
   \leq {\hat{h}_{\Phi_1}}(z, h(x, u), u) ^\top {\hat{h}_{\Phi_2}}(z, h(x, u), u)
   \end{aligned}
  \end{multline*}
   for all $x\in \mathcal{X}$, $z\in \mathcal{Z}$ and $u\in\mathcal{U}$.
 
\subsection{Examples of Dynamic Dissipativity}
  
Several motivating examples are provided in this subsection. The first is adopted from \cite{Lanzon23}, which provides a class of negative imaginary systems characterised by an LTI auxiliary system and a dynamic supply rate. The example is paraphrased in terms of Definition~\ref{def: dynamic_dissi}.

\begin{exmp}[Input-output negative imaginariness]\label{exmp: IONI} 
Let $\Sigma$ be a stable LTI system with a minimal realisation $(A, B, C, D)$ in which $D = D^\top$. Then, $\Sigma(s) = C(sI - A)^{-1}B + D$ is said to be IONI with a level of output strictness $\delta \ge 0$ and a level of input strictness $\epsilon \ge 0$ having an arrival rate $\alpha \in \mathbb{N}$ and departure rate $\beta \in \mathbb{N}$ (i.e.~${\rm IONI}_{(\delta, \epsilon, \alpha, \beta)}$)~\cite[Def.~1]{Lanzon23} if
\begin{align*}
& j\omega [\Sigma(j\omega) - \Sigma(j\omega)^*] - \delta \omega^2 \bar{\Sigma}(j\omega)^* \bar{\Sigma}(j\omega) \\
& \qquad - \epsilon \left( \frac{\omega^{2\beta}}{1 + \omega^{2(\alpha + \beta - 1)}} \right) I \ge 0 \quad \forall \omega \in \Real \cup \{\infty\},
\end{align*}
where $\bar{\Sigma}(j\omega) = \Sigma(j\omega) - D$. Moreover, $\Sigma$ is said to be output strictly negative imaginary (OSNI) if it is ${\rm IONI}_{(\delta, \epsilon, \alpha, \beta)}$ with $\delta > 0$, $\epsilon \ge 0$, $\alpha, \beta \in \mathbb{N}$ and $\Sigma(s) - \Sigma(-s)^\top$ has full normal rank~\cite[Def.~7]{Lanzon23}. Let $\Phi$ be a stable LTI auxiliary system with state $z$ for which $z(0) = 0$, minimal realisation $(A_\Phi, B_\Phi, C_\Phi, D_\Phi)$, and transfer function $\Phi(s) = C_\Phi(sI - A_\Phi)^{-1}B_\Phi + D_\Phi$ satisfying
\[
\Phi(-s)\Phi(s) = \frac{(-s)^\beta s^{\beta}}{1+(-s)^{(\alpha+\beta-1)}s^{(\alpha+\beta-1)}};
\]
see~\cite[Lem.~1]{Lanzon23}. For instance, in the special case of $\alpha=\beta=1$, the auxiliary system $\Phi(s)=\frac{s}{s+1}$ has a realisation  $(-1, 1, -1, 1)$. By~\cite[Th.~4]{Lanzon23} and Lemma~\ref{prop: differential_dissi}, it may be verified that an ${\rm IONI}_{(\delta, \epsilon, \alpha, \beta)}$ $\Sigma$ is $\Xi$-dissipative with the dynamic supply rate
\begin{equation}\label{eq: IONI_supply}
   \Xi(u, y, \xbar)(t)  = 2\dot{\bar{y}}(t)^\top u(t) - \delta \dot{\bar{y}}(t)^\top \dot{\bar{y}}(t)  - \epsilon  (\Phi u)(t)^\top (\Phi u)(t) 
\end{equation}
and storage function of the form
\[
S(x, z) = \TwoOne{x}{z}^\top P \TwoOne{x}{z},
\]
where $P = P^\top$. Note that the operator $\Xi$ in \eqref{eq: IONI_supply} involves not only the auxiliary system $\Phi$ but also the differentiation operation on $\bar{y}$, i.e. $\dot{\bar{y}}$. Furthermore, by \cite[Cor.~8]{Lanzon23} and Lemma~\ref{prop: differential_dissi}, it may be verified that an OSNI $\Sigma$ is $\Xi$-dissipative with the supply rate
\[
\Xi(u, y, \xbar)(t)  = 2\dot{\bar{y}}(t)^\top u(t) - \delta \dot{\bar{y}}(t)^\top \dot{\bar{y}}(t)
\]
and storage function $S(x) = x^\top P x$ with $P = P^\top$. IONI and OSNI systems are examples in which the dynamics of the auxiliary system $\Phi$ are not necessarily those of the supply rate $\Xi$.
\end{exmp}
 
Next, we provide a couple of examples of nonlinear systems that are dissipative with respect to dynamic supply rates. They may serve the purpose of illustrating the
search for auxiliary systems in order to satisfy certain desired dissipation inequalities. The following example provides a $\Xi'$-dissipative system $\Sigma$ whose supply rate $\Xi$ depends on the initial condition $x(0)$ of $\Sigma$.

\begin{exmp}\label{exmp: ICD}
Let $\Sigma$ in \eqref{eq: OLSystem} be described by
\begin{equation}\label{eq: numerical_exmp_ICD} 
  \Sigma:\;
  \begin{aligned}
\dot{x}_{1}(t) &= - a x_{1}(t)  -\psi(x_1(t)) +2x_2(t) \\  & \qquad -\textstyle\sum_{k=-N}^{M} b_k (x_1(t))^{2k+1} 
+u(t)\\ 
\dot{x}_{2}(t)  &= -x_2(t) +u(t)\\
y(t)&= x_1(t)-x_{2}(t)
  \end{aligned}
\end{equation}
with $x(0) = \STwoOne{x_1(0)}{x_2(0)} \in \Real^2$, where $a\geq 1$, $b_k\geq 0$ for all $k \in \{-N, -N+1, \ldots,  M\}$, $N$ and $M$ are nonnegative integers, and $\psi: \mathbb{R}\rightarrow\mathbb{R}$ is a locally integrable nonlinearity satisfying $\psi(0)=0$ and ${\psi}(r) r \geq 0$ for all $r\in\mathbb{R}$. Let the dynamic supply rate $\Xi=\SThreeOne{u}{y}{x(0)} \mapsto \xi$ be given by
\begin{equation}\label{eq: exmple_ICD_supply}
   \Xi  :\;\begin{aligned}
\dot{z}(t) &= -  z(t)  +  u(t), \quad z(0) = \SOneTwo{0}{1} x(0) \\ 
\xi(t)&=  u(t)(3z(t)+y(t)).
  \end{aligned}
\end{equation} 
Consider the candidate storage function $S(x_1, x_2, z)=\frac{1}{2} x_1^{2}+ \frac{1}{2}x_2^2+ \frac{1}{2}z^2$. In view of \eqref{eq: exmple_ICD_supply}, let the auxiliary system $\Phi$ have dynamics $\dot{z}(t) = -z(t) + u(t)$ with $z(0)=x_2(0)$, i.e. the same as the state equation in \eqref{eq: exmple_ICD_supply}. Observe that $x_2(t)=z(t)\;\forall t\geq 0$. Along the solutions to $\Sigma$ and $\Phi$,  
\begin{align}
  &\frac{d}{dt}S(x_1, x_2, z)\notag\\
 =& -ax_1^2 +2x_1x_2 -x_1\psi(x_1) - \textstyle \sum_{k=-N}^M b_k (x_1)^{2k+2}  + ux_1 \notag\\
    &- x_2^2 +ux_2 + z(-z+u) \notag\\
     \leq& -(a-1)x_1^2- (x_1^2 -2x_1x_2+x_2^2) -z^2 + u(x_1+x_2+z)\notag\\
     \leq&~u(x_1+x_2+z) -y^2 \leq u(x_1+2x_2) = \xi. \label{eq: ICD_example}
\end{align}
Integrating both sides gives that $\Sigma$ is  $\Xi'$-dissipative by Lemma~\ref{prop: differential_dissi}.
\end{exmp}

Example~\ref{exmp: ICD} will be revisited in Section~\ref{sec: simu} where a simulation example is provided. The next two nonlinear examples concern $\Xi$-dissipativity.

\begin{exmp}\label{ex: static}
  Let $\Sigma$  be defined by $y(t)={\sigma}(u(t))$, where
  $\sigma: \mathbb{R}\rightarrow\mathbb{R}$ is locally integrable and satisfies ${\sigma}(0)=0$ and
\begin{align}
  {\sigma}(r)(br - {\sigma}(r)) &\geq 0 \; \forall r \in \mathbb{R},\label{eq: sector_bounded} \\
( {\sigma}(r)- {\sigma}(q))(r-q)&\geq 0 \; \forall r, q \in \mathbb{R},\label{eq: monotone}
\end{align}
with $b>0$; i.e., $\Sigma$ is sector-bounded in the sector $[0, b]$ and monotonically nondecreasing.  Adding
\eqref{eq: sector_bounded} to \eqref{eq: monotone} and rearranging the inequality yield 
 \begin{equation}\label{eq: examp_1}
  \begin{aligned}
  (1+b){\sigma}(r)r-({\sigma}(r))^2   -{\sigma}(r)q \geq - {\sigma}(q)q  + {\sigma}(q)r 
 \end{aligned} 
 \end{equation}
 for  $r, q\in \Real$. Let the supply rate $\Xi=\SThreeOne{u}{y}{\xbar} \mapsto \xi$ be given by
\begin{equation*} 
    \Xi :\; 
  \begin{aligned}
    \dot{z}(t)  &= -z(t) + u(t),\; z(0)=z_0\in \Real\\
   \xi(t)  &=  -y(t)\left[z(t)-(1+b)u(t)+y(t)\right].
  \end{aligned} 
\end{equation*} 
In view of the staticity of $\Sigma$, consider the candidate storage function $S(z)=\int_{0}^{z}{\sigma}(r)\, dr$. Note that $S$ belongs to $C^1$ and
 $\frac{d}{dt}S(z)={\sigma}(z)\dot{z}$. In light of \eqref{eq: examp_1}, let the auxiliary system $\Phi$ have dynamics
 $\dot{z}(t) = -z(t) + u(t)$ with $z(0)=z_0$. Then, along the solutions to $\Phi$ and by using \eqref{eq: examp_1}, we have
 $\frac{d}{dt}S(z) \leq (1+b)y u - y^2 -yz=\xi$. 
Integrating both sides yields that $\Sigma$ is $\Xi$-dissipative by Lemma~\ref{prop: differential_dissi}.
\end{exmp}
 
\begin{exmp}\label{ex: dynamic}
 Let $\Sigma$ in \eqref{eq: OLSystem}  be  given by
  \begin{equation*} 
    \Sigma :\;
    \begin{aligned}
      \dot{x}_{1}(t)  &= x_{2} \\
      \dot{x}_{2}(t) &= - (x_{1}(t))^3 + ({\psi}(x_{2}(t)))^2 + u(t)\\
      y(t)&= x_{2}(t)
    \end{aligned}
  \end{equation*}
  with $x(0)=\STwoOne{x_{1,0}}{x_{2,0}}$, where ${\psi}: \mathbb{R}\rightarrow\mathbb{R}$ is locally integrable and satisfies ${\psi}(0)=0$ and $ {\psi}(r) r \geq 0$ for all $r\in\mathbb{R}$. Let the dynamic supply rate $\Xi=\SThreeOne{u}{y}{\xbar} \mapsto \xi$ be given by  
  \begin{equation*} 
   \Xi :\;
     \begin{aligned}
      \dot{z}(t)  &= -z(t) + {\psi}(z(t))(u(t))^2 + y(t),\ z(0)=z_0\in \Real\\
     \xi(t)  &=  y(t)\left[z(t) + u(t)+ ({\psi}(y(t)))^2\right].
    \end{aligned} 
  \end{equation*}
  Consider the candidate storage function $S(x_1, x_2, z)=\frac{1}{4}x_{1}^4 + \frac{1}{2}x_{2}^2 +\frac{1}{2}z^2$. Then, we obtain  
    $\frac{d}{dt}S(x_1, x_2, z) =y({\psi}(y))^2 + y u + z \dot{z}$. 
Note that ${\psi}(z(t))z(t)\geq 0$ for all $z(t) \in \mathbb{R}$ and thus   
  $z(t)\left[ -z(t) -{\psi}(z(t))(u(t))^2 \right] \leq 0$
  for all $z(t) \in \mathbb{R}$. Using the auxiliary system $\Phi$ with dynamics $\dot{z}(t) = -z(t) + {\psi}(z(t))(u(t))^2 +
y(t)$ with $z(0)=z_0$, it then follows that
\begin{equation*} 
  \begin{aligned}
  \frac{d}{dt}S(x_1, x_2, z)
  \leq  y({\psi}(y))^2 + y u  + z y =\xi. 
  \end{aligned}
  \end{equation*}
 By Lemma~\ref{prop: differential_dissi}, $\Sigma$ is $\Xi$-dissipative.
  Alternatively, one may choose an empty $\Phi$ with $\mathcal{Z}=\emptyset$. In such a case, a viable supply rate is
  $\tilde{\xi}(t)=\tilde{\Xi}(u(t), y(t))=u(t)y(t) + y(t)(\phi(y(t)))^2$, which is static.
\end{exmp}

In both Examples~\ref{ex: static} and \ref{ex: dynamic}, even though the systems may be described by dissipativity with respect to some static supply rates, the advantage of using dynamic supply rates lies in offering great flexibility in system characterisation as well as reducing conservatism in feedback stability analysis, similarly to the benefit of using dynamic multipliers, or IQCs, in an input-output setting.

\section{Main Results on Feedback Stability} \label{sec: FB}

In this section, we present the main results of this paper --- feedback stability analysis via the proposed dissipativity with dynamic supply
rates. The results involve Lyapunov and  asymptotic stability presented in the order stated. First, some technical definitions are stated and the
configuration under study is introduced.

\subsection{Definitions}

Consider the system $\Sigma$ described in \eqref{eq: OLSystem}.

  \begin{defn}
    The system $\Sigma$ in \eqref{eq: OLSystem} is said to be \emph{zero-state detectable} if $u(t) = 0$  and $y(t) = 0$ for all $t \geq 0$ implies
    $\displaystyle\lim_{t \to \infty} x(t) = 0$.
  \end{defn}
  
  \begin{defn}
  Let $x(t)$ be a solution of $\Sigma$ in \eqref{eq: OLSystem} with $u = 0$. A point $p \in \mathcal{X}$ is said to be a \emph{positive limit point} of $x(t)$ if there exists a
  sequence $\{t_n\}_{n=1}^{\infty}$, with $t_n \to \infty$ as $n \to \infty$, such that $x(t_n) \to p$ as $n \to \infty$. The set of all positive limit points of $x(t)$
  is called the \emph{positive limit set} of $x(t)$.
  \end{defn}
  
  \begin{defn}
    A set $M \subset \mathcal{X}$ is said to be a \emph{positively invariant set} with respect to $\Sigma$ in \eqref{eq: OLSystem} if
    \[
     u = 0 \quad \text{and} \quad x(0) \in M \implies x(t) \in M \quad \forall t \geq 0.
     \]
  \end{defn}

  \begin{defn}\label{def:stability}
      $\Sigma$ in \eqref{eq: OLSystem} with $u = 0$ is said to be:
    \begin{enumerate} \renewcommand{\theenumi}{\textup{(\roman{enumi})}}\renewcommand{\labelenumi}{\theenumi}
    \item \textit{Lyapunov stable with respect to $x$} if, for every $\epsilon>0$, there exists $\delta=\delta(\epsilon)>0$ such
      that $\norm{x(0)}<\delta$ implies that $\norm{x(t)}<\epsilon$ for all $t\geq 0$;
  
    \item \textit{asymptotically stable with respect to $x$} if it is Lyapunov stable with respect to $x$   and
      there exists $\delta>0$ such that $\norm{x(0)}<\delta$ implies that $\displaystyle\lim_{t\to \infty} x(t)=0$.
  
     \item \textit{globally asymptotically stable with respect to $x$} if it is Lyapunov stable with respect to $x$  and $\displaystyle\lim_{t\to \infty} x(t)=0$ for all $x(0) \in \mathcal{X}$.
    \end{enumerate}
  \end{defn}

\subsection{Problem Formulation and Preliminaries}

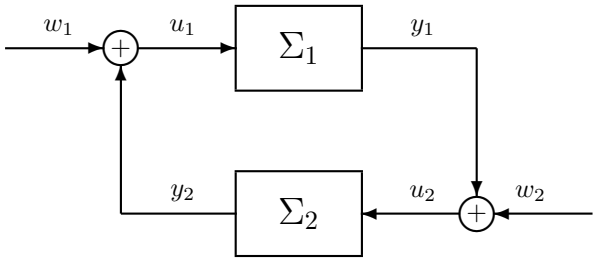
\begin{figure}[htb]
  \centering
  \setlength{\unitlength}{1.1mm}
  \begin{picture}(65,34)
  \thicklines \put(-4,25){\vector(1,0){12}} \put(10,25){\circle{4}}
  \put(10,25){\makebox(0,0){$+$}}
  \put(12,25){\vector(1,0){12}} \put(24,20){\framebox(15,10){\large$\Sigma_1$}}
  \put(39,25){\line(1,0){14}} \put(53,25){\vector(0,-1){18}}
  \put(51,5){\vector(-1,0){12}} \put(53,5){\circle{4}}  \put(53,5){\makebox(0,0){$+$}}
  \put(67,5){\vector(-1,0){12}} \put(24,0){\framebox(15,10){\large$\Sigma_2$}}
  \put(24,5){\line(-1,0){14}} \put(10,5){\vector(0,1){18}}
  \put(15,5){\makebox(5,5){$y_2$}} \put(44,25){\makebox(5,5){$y_1$}}
  \put(0,25){\makebox(5,5){$w_1$}} \put(57,5){\makebox(5,5){$w_2$}}
  \put(15,25){\makebox(5,5){$u_1$}} \put(44,5){\makebox(5,5){$u_2$}}
  \end{picture}\caption{Feedback configuration of $\Sigma_1$ and $\Sigma_2$.} \label{fig: feedback}
  \end{figure}

Consider two nonlinear input-state-output systems
\begin{equation}\label{eq: OLSystems}
  \Sigma_i:\;
  \begin{aligned}
    \dot{x}_i(t) &= f_i(x_i(t), u_i(t)), \; x_i(t) \in \mathcal{X}_i, \; u_i(t) \in \mathcal{U}_i, \\
    y_i(t) &= h_i(x_i(t), u_i(t)), \; y_i(t) \in \mathcal{Y}_i
  \end{aligned}
\end{equation}
for $i\in \{ 1, 2\}$,  with $x_i(0)=x_{i,0}$, $\mathcal{X}_i=\Real^{n_i}$, $\mathcal{U}_1 = \mathcal{Y}_2 = \Real^m$, $\mathcal{U}_2 = \mathcal{Y}_1 = \Real^p$,  locally Lipschitz $f_i:\mathcal{X}_i\times\mathcal{U}_i\to \mathcal{X}_i$ and continuous $h_i:\mathcal{X}_i\times \mathcal{U}_i\to \mathcal{Y}_i$  interconnected in a feedback configuration shown in  Fig.~\ref{fig: feedback} and described by
\begin{align} \label{eq: FB} u_1 =  w_1 + y_2; \quad u_2 = w_2 + y_1.
\end{align}
Denote the admissible inputs set by $\mathscr{U}_i$ and the outputs set over $\mathscr{U}_i$ by $\mathscr{Y}_i$ for $i\in \{ 1, 2\}$.

Henceforth, the feedback interconnection  of Fig.~\ref{fig: feedback} is denoted by
$\Sigma_1 \| \Sigma_2$ and written as
\begin{equation} \label{eq: FB_state}
  \hspace{-2.3mm} \Sigma_1 \| \Sigma_2:\;
  \begin{aligned}
    \dot{x}(t) & = \check{f}(x(t), w(t)), \; x(t) \in \mathcal{X},  w(t)\in \mathcal{W},  \\
    y(t) & =  \check{h}(x(t),  w(t)), \; y(t)\in \mathcal{Y},
  \end{aligned}
\end{equation}
where $x(0)=(x_{1,0},x_{2,0})$, $\mathcal{X} = \mathcal{X}_1 \times
\mathcal{X}_2$, ${\mathcal{W}} = \mathcal{U}_1 \times
\mathcal{U}_2$, $\mathcal{Y} = \mathcal{Y}_1 \times \mathcal{Y}_2$, $ x(t) = (x_1(t), x_2(t))$, {$w(t) = (w_1(t), w_2(t))$} and $y(t) = (y_1(t), y_2(t))$. 

 We associate an auxiliary system  $\Phi_i$ of the form $\eqref{eq: aux}$ with state variable $z_i \in \mathcal{Z}_i=\Real^{n_{z_i}}$ for
$i\in \{1, 2\}$  with each system $\Sigma_i$. That is, 
\begin{equation*} 
\Phi_i:\; 
\begin{aligned}
  \dot{z}_i(t)& = g_i(z_i(t), x_i(t), u_i(t)), \; z_i(t) \in \mathcal{Z}_i, z_i(0)=z_{i,0}\\
  \phi_i(t) &  =  h_{\Phi_i}(z_i(t), x_i(t), u_i(t)),\; \phi_i(t) \in \mathcal{O}_i.
\end{aligned}
\end{equation*}
In what follows, let $\mathcal{Z} = \mathcal{Z}_1 \times \mathcal{Z}_2$, $\mathcal{O}=\mathcal{O}_1 \times \mathcal{O}_2$,  
$z(t) = (z_1(t), z_2(t))$ and $\phi(t)=(\phi_1(t), \phi_2(t))$. The two auxiliary systems to the feedback system $\Sigma_1 \| \Sigma_2$ may thus be described succinctly as
\begin{equation}\label{eq: two_aux}
  \begin{aligned}
    \dot{z}(t) &= \check{g}(z(t), x(t), w(t)),\; z(t) \in \mathcal{Z},  z(0)= (z_{1,0}, z_{2,0})\\
    \phi(t) & =  \check{h}_{\Phi}(z(t), x(t), w(t)),\; \phi(t) \in \mathcal{O}.
\end{aligned}
\end{equation}

Define the operator $\Gamma: \mathscr{U}\times \mathscr{Y} \times  {\bar{\mathcal{X}}} \to \mathscr{Y}\times \mathscr{U} \times {\Xbar}$ by  $\Gamma(u, y, {\xbar})=(y, u, {\xbar})$, i.e.  $\Gamma$ swaps the order
of its first two arguments. Denote by $\circ$ the composition operation.

\subsection{Lyapunov Stability}

Main results on feedback Lyapunov stability are derived in this section. First, an assumption on the storage functions is stated. 
\begin{assum}\label{assum:bounded_S}
  The storage functions $S_1$ and $S_2$ are $C^1$ and there exist $\delta > 0$ and class
  $\mathcal{K}$ functions $\alpha$ and $\beta$ such that
  \begin{equation}\label{eq:partial_bounded_S}
    \begin{aligned}
      \alpha(\norm{x}) \leq  \sum_{i = 1}^2 S_i(x_i, z_i) \leq \beta(\norm{x})
    \end{aligned} 
  \end{equation}
  for all  $x\in \mathcal{X}$ with $\norm{x} < \delta$ and $z \in \mathcal{Z}$. 
\end{assum}
 
 Assumption~\ref{assum:bounded_S} requires boundedness on the sum of two storage functions in terms of parts (but not all) of their arguments. This resembles boundedness on a time-varying Lyapunov candidate function~\cite[Th.~4.8]{Kha02} and a Lyapunov candidate function for partial stability~\cite[Th.~4.1]{Haddad08}. An alternative assumption where the lower and upper bounds are functions of $\|z\|$ will be considered later.

\begin{thm} \label{thm: Lya_sta}
  Suppose there exists a causal operator $\Xi: \mathscr{U}_1 \times \mathscr{Y}_1 \times \mathcal{X}_1 \to \boldsymbol{\rm L}_{1e}$ such that $\Sigma_1$ is $\Xi'$-dissipative and $\Sigma_2$ is $(-\Xi\circ \Gamma)$-dissipative. Furthermore, suppose the corresponding storage functions $S_1$ and $S_2$ satisfy Assumption~\ref{assum:bounded_S}. Then, $x^* = (x_1^*, x_2^*) = (0, 0)$ is a Lyapunov stable equilibrium  of the
    closed-loop system $\Sigma_1 \| \Sigma_2$ with $w_1 = 0$ and $w_2 = 0$.  If additionally $\Sigma_2$ is static, then $\mathcal{X}_2=\emptyset$ and $x_1^* = 0$ is a Lyapunov stable equilibrium.
\end{thm}
 
\pf
By hypothesis and using the time-invariant properties of $\Sigma_1$ and $\Sigma_2$,  we have
  \begin{align}\label{eq: diss_ineq_Lya_sta}
   \hspace{-2mm} \begin{split}
      S_1(x_1(t_2), z_1(t_2))  & \leq  S_1(x_1(t_1), z_1(t_1))  \\
      & \qquad + \int_{t_1}^{t_2} \Xi(u_1, y_1, {x_1(t_1)})(t) \, dt \\
      S_2(x_2(t_2), z_2(t_2)) & \leq S_2(x_2(t_1), z_2(t_1))  \\
      & \qquad - \int_{t_1}^{t_2}  \Xi(y_2, u_2, {x_1(t_1)})(t) \, dt
    \end{split}
  \end{align}
  for all $t_2\geq t_1$, all initial conditions $x_1(t_1)\in\mathcal{X}_1, x_2(t_1)\in\mathcal{X}_2$, and all input functions $u_1 \in \mathscr{U}_1$,
  $u_2 \in  \mathscr{U}_2$. Substituting the feedback equations \eqref{eq: FB} with $w_1 = 0$, $w_2 = 0$ into the inequalities in \eqref{eq: diss_ineq_Lya_sta}, summing
  them,  dividing both sides by $t_2-t_1$, and taking $t_2 \to t_1$ then yields
  \begin{align} \label{eq: S_dot} \frac{d}{dt} (S_1(x_1, z_1) + S_2(x_2, z_2)) \leq 0
  \end{align}
  along the solutions to \eqref{eq: FB_state} and \eqref{eq: two_aux}. Define $V(x, z) = S_1(x_1, z_1) + S_2(x_2, z_2)$. We have from \eqref{eq: S_dot} that
  \[
    \frac{d}{dt} V(x, z) = \dot{V}(x, z) \leq 0
  \]
  along the solutions to \eqref{eq: FB_state} and \eqref{eq: two_aux}. This implies that $V$ is nonincreasing along the solutions to \eqref{eq:
    FB_state} and \eqref{eq: two_aux}. From (\ref{eq:partial_bounded_S}), we have $V(x^*, z) = 0$ and $V(x, z) > 0$ for all $x\in \mathcal{X}$ with
  $\norm{x} < \delta$, $x \neq x^*$, and $z\in \mathcal{Z}$. This means $(x^*, z)$ is a strict minimum of $V$ for all $z\in \mathcal{Z}$, whereby
   $\check{f}(x^*, 0) = 0$, i.e.  $x^* = 0$ is an equilibrium in \eqref{eq: FB_state} with $w_1 = 0$ and $w_2 = 0$. That $x^*=0$ is Lyapunov stable then follows from
  \cite[Th.~4.1(ii)]{Haddad08}.  For the case when $\Sigma_2$ is additionally static, $\mathcal{X}_2=\emptyset$ and we have
  \[
    S_2(z_2(t_2)) \leq S_2(z_2(t_1)) - \int_{t_1}^{t_2} \Xi(y_2, u_2, \bar{x})(t) \, dt
  \]
  for all $t_2\geq t_1$, $\bar{x} \in \Xbar$, $u_2 \in \mathscr{U}_2$, and $y_2(t) = \Sigma_2(u_2(t))$. With $V(x_1, z) = S_1(x_1, z_1) + S_2(z_2)$, the arguments above may be repeated to show that $x_1^*=0$ is Lyapunov stable. \hfill $\square$
\endpf

The purpose of the operator $\Gamma$ in Theorem~\ref{thm: Lya_sta} is to represent the ``\emph{inverse supply rate}'' for $\Sigma_2$, namely,
$(\Xi \circ \Gamma) (u_2, y_2, \xbar) = \Xi(y_2, u_2, \xbar)$. Theorem~\ref{thm: Lya_sta} indicates the main idea of this paper: \emph{If $\Sigma_1$ and $\Sigma_2$ can be simultaneously described by ``complementary'' dynamic supply rates $\Xi'$ and $-\Xi \circ \Gamma$, then the closed-loop system $\Sigma_1 \| \Sigma_2$ will have
Lyapunov stability with respect to $x$.} This generalises the classical idea of feedback stability via static dissipativity, and the connection is
elaborated in Section~\ref{sec: examples}.  It is noteworthy that $\Sigma_1$ and $\Sigma_2$ may be associated with different auxiliary systems  $\Phi_1$ and $\Phi_2$ with
state variables $z_1$ and $z_2$, respectively, as long as the dissipativity conditions on $\Sigma_1$ and $\Sigma_2$ hold. More importantly, the
dynamics involving $z_1$ and $z_2$ are not necessarily stable since their stability is irrelevant as far as closed-loop stability is concerned.

 Next, we present a feedback Lyapunov stability result under a different assumption on the storage functions that may be more useful in certain circumstances than Assumption~\ref{assum:bounded_S}. 

\begin{assum}\label{assum:bounded_stable_S}
The storage functions $S_1$ and $S_2$ are $C^1$ and there exist $\delta > 0$ and class
  $\mathcal{K}$ functions $\alpha$ and $\beta$ such that
  \begin{equation}\label{eq:bounded_stable_S}
    \begin{aligned}
      \alpha(\norm{(x, z)}) \leq  \sum_{i = 1}^2 S_i(x_i, z_i) \leq \beta(\norm{(x, z)})
    \end{aligned} 
  \end{equation}
  for all   $x\in \mathcal{X}$ and $z \in \mathcal{Z}$ with $\norm{(x, z)} < \delta$. 
\end{assum}
 In contrast to Assumption~\ref{assum:bounded_S}, the lower and upper bounds in Assumption~\ref{assum:bounded_stable_S} depend on both $x$ and $z$. A byproduct of this assumption is that the stability with respect to both $x = 0$ and $z = 0$ in \eqref{eq: FB_state} and \eqref{eq: two_aux} may be established, even though we are only concerned with the former. This resembles the theory of dynamic Lyapunov functions proposed in \cite[Def.~1]{Sassano:13}.

\begin{thm} \label{thm: Lya_sta_2}
 Suppose there exists a causal operator $\Xi:\mathscr{U}_1 \times \mathscr{Y}_1 \times \Xbar \to \boldsymbol{\rm L}_{1e}$ such that $\Sigma_1$ is  $\Xi'$-dissipative and $\Sigma_2$ is $(-\Xi\circ \Gamma)$-dissipative. Furthermore, suppose the corresponding storage functions $S_1$ and $S_2$ satisfy Assumption~\ref{assum:bounded_stable_S} and $(x^*, z^*) = (x_1^*, x_2^*, z_1^*, z_2^*) = (0, 0, 0, 0)$ is an equilibrium of \eqref{eq: FB_state} and \eqref{eq: two_aux} with $w_1=0$ and $w_2=0$. Then, the equilibrium $x^*=(0, 0)$  of   $\Sigma_1 \| \Sigma_2$ in \eqref{eq: FB_state} is Lyapunov stable. If additionally $\Sigma_2$ is static, then $\mathcal{X}_2=\emptyset$ and the equilibrium $x_1^* = 0$ is Lyapunov stable.
\end{thm}
\pf
Define $V(x, z) = S_1(x_1, z_1) + S_2(x_2, z_2)$.  Following the same arguments  in the proof of Theorem~\ref{thm: Lya_sta}, we can obtain that 
   $\frac{d}{dt} V(x, z) = \dot{V}(x, z) \leq 0$
  along the solutions to \eqref{eq: FB_state} and \eqref{eq: two_aux}. By the  standard Lyapunov stability theorem \cite[Th. 3.2.4]{Sch17},   the equilibrium $(x^*, z^*)$ of  \eqref{eq: FB_state} and \eqref{eq: two_aux} is Lyapunov stable. By \cite[Lem. 4.5]{Kha02}, this is equivalent to the existence of $c > 0$ and class $\mathcal{K}$ function $\kappa$ such that
  \[
  \|(x(t), z(t))\| \le \kappa(\|(x(0), z(0))\|)
  \]
  for all $t \ge 0$ and all $(x(0), z(0))$ satisfying $\|(x(0), z(0))\| < c$. This implies
  \[
  \|x(t)\| \le \|(x(t), z(t))\| \le \kappa(\|(x(0), {0})\|) = \kappa(\|x(0)\|)
  \]
  for all $t \ge 0$ and $\|x(0)\| < c$, from which Lyapunov stability of $x^* = (0, 0)$ of \eqref{eq: FB_state} follows again by \cite[Lem. 4.5]{Kha02}. The static case can be similarly proved by using $V(x_1, z) = S_1(x_1, z_1) + S_2(z_2)$. \hfill $\square$
\endpf

 \begin{rem}\label{rem: parallel_result}
  In Theorem~\ref{thm: Lya_sta_2},  $(x^*, z^*)$ is presumed to be an equilibrium. By contrast,  Theorem~\ref{thm: Lya_sta} establishes that $x^*$ is an equilibrium using properties of the storage functions in Assumption~\ref{assum:bounded_S}. 
\end{rem}

Feedback stability in the sense of Lyapunov often leaves much to be desired. Next, we examine the stronger notion of asymptotic feedback stability via dissipativity.

\subsection{Asymptotic Stability}
In this subsection, we establish feedback asymptotic stability via dissipativity with dynamic supply rates. The following technical lemma
is needed  in the proof of Theorem~\ref{thm: diss}. It mimics \cite[Prop.~3.2.16]{Sch17} and establishes asymptotic stability  for an open-loop system through dissipativity.

\begin{lem}\label{prop:dissi_stability}
  Let $\Sigma$ in \eqref{eq: OLSystem} be zero-state detectable and $\Xi$-dissipative on $(\mathcal{X}, \mathscr{U})$ with an auxiliary system  $\Phi$ given in \eqref{eq: aux},  and a \emph{static} supply rate $\tilde{\Xi}(u(t), y(t))$ that is continuous in $y(t) \in \mathcal{Y}$ and satisfies $\tilde{\Xi}(0, y(t)) \leq 0$ for all $y(t) \in \mathcal{Y}$. Let $(0, 0) \in (\mathcal{X}, \mathscr{U})$ and $\tilde{\Xi}(0, y(t)) = 0$ imply $y(t)=0$. Suppose also that the corresponding storage function $S(x, z)$ is
  $C^1$ and there exist $\delta > 0$ and class $\mathcal{K}$ functions $\alpha, \beta$ such that
  \begin{align} \label{eq: S_bounds}
    \alpha(\norm{x})\leq S(x, z)\leq \beta(\norm{x})
  \end{align}
  for all  $x\in \mathcal{X}$ with $\norm{x} < \delta$ and $z\in \mathcal{Z}$. Then, $x^*=0$ is an asymptotically stable equilibrium of  $\Sigma$ in \eqref{eq:
    OLSystem} with $u = 0$. If, additionally, $\alpha, \beta$ are class $\mathcal{K}_\infty$ functions and \eqref{eq: S_bounds} holds for all
  $x \in \mathcal{X}$ and $z \in \mathcal{Z}$, then $x^*=0$ is globally asymptotically stable.
\end{lem}

\pf
  By hypothesis and using the time-invariant property of $\Sigma$, with $u = 0$,  we have the following dissipation inequality
  \[
    S(x(t_2), z(t_2)) \leq S(x(t_1), z(t_1)) + \int_{t_1}^{t_2} \tilde{\Xi}(0, y(t)) \, dt.
  \]
  Dividing both sides by $t_2-t_1$ and taking  the limit as $t_2\to t_1$,  we get
  \begin{align*}
    \frac{d}{dt}S(x(t), z(t)) \leq \tilde{\Xi}(0, y(t)) \leq 0, 
  \end{align*}
  and thus $S(x, z)$ is nonincreasing along the solutions of \eqref{eq: OLSystem} with $u=0$.  Since $S(x, z)$ has a strict local minimum in $x$
  at $x^* = 0$ and $S(x^*, z) = 0$ for all $z \in \mathcal{Z}$ according to \eqref{eq: S_bounds}, it follows that $x^* = 0$ is an equilibrium, i.e.
  $f(x^*, 0) = 0$. Note that $\beta^{-1}\circ\alpha$ is a class $\mathcal{K}$ function \cite[Lem.~4.2]{Kha02}. Given any $\epsilon>0$, let
  $\delta=\delta(\epsilon)$ be chosen such that $\delta=\beta^{-1}(\alpha(\epsilon))>0$.  Then by hypothesis  and since $S(x, z)$ is nonincreasing along the solutions of \eqref{eq: OLSystem} with $u=0$, for $\norm{x(0)}<\delta$, we
  have
  \begin{align*}
    \alpha(\norm{x(t)})\leq S(x(t), z(t))\leq S(x(0), z(0)) \leq \beta(\norm{x(0)})<\beta(\delta)
  \end{align*}
  for all $t> 0$. This implies $\norm{x(t)} < \alpha^{-1}(\beta(\delta)) = \epsilon$ for all $t> 0$. Then $x^*=0$ is a Lyapunov stable equilibrium.  

  To further show asymptotic stability of $x^*=0$, let $W(x) = -\tilde{\Xi}(0, h(x, 0))$. Since $W$ is continuous, $W(x) \geq 0$ and
  $W(0) = 0$, by \cite[Th.~4.2]{Haddad08}, the trajectory $x(t)$ with $\norm{x(0)} < \delta$ approaches the set
  $C =\left\{ x\in \mathcal{X}~|~\tilde{\Xi}(0, h(x, 0))=0\right\}$ as $t\to\infty$ and thus the positive limit set of $x(t)$ is a subset of the set
  $C$. Furthermore, by~\cite[Lem.~4.1]{Kha02}, the positive limit set of $x(t)$ is a positively invariant set. Suppose $x(t)$ as a solution to
  \eqref{eq: OLSystem} stays identically for all $t\geq 0$ in the set $C$. Since $\tilde{\Xi}(0, h(x(t), 0))=0$ implies $y(t) = h(x(t), 0)=0$ for all
  $t\geq 0$ and $\Sigma$ is zero-state detectable, we have $\displaystyle\lim_{t\to \infty} x(t)=0$. In other words, no solution other than
  $x^*=0$ stays identically in the set $C$ for all $t \geq 0$. Asymptotic stability of $x^* = 0$ then follows.

  If, additionally, $\alpha, \beta$ are class $\mathcal{K}_\infty$ functions and \eqref{eq: S_bounds} holds for all $x \in \mathcal{X}$ and
  $z \in \mathcal{Z}$, global asymptotic stability of $x^* = 0$  can be concluded by using the same arguments as above. \hfill $\square$
  \endpf

A strict version of dynamic dissipativity beyond Definition~\ref{def: dynamic_dissi} is given below. It is needed to establish asymptotic closed-loop stability.
 
\begin{defn}[I/O-strict dynamic dissipativity]\label{def: dynamic_strict_dissi}
 Let $\Xi: \mathscr{U} \times \mathscr{Y} \times \Xbar \to \boldsymbol{\rm L}_{1e}$ be causal, and $\gamma_1$ and $\gamma_2$ be locally integrable  continuous
  positive definite functions. $\Sigma$ in \eqref{eq: OLSystem} is called \emph{very strictly $\Xi$-dissipative}  if  there exist an auxiliary system~\eqref{eq: aux} and  $S: \mathcal{X} \times \mathcal{Z} \to \Real$ such that 
  \begin{equation}\label{eq: strict_dissipation}
    \begin{aligned}
    S(x(T), z(T)) + &\int_{0}^{T}  [\gamma_1(u(t)) +  \gamma_2(y(t))] \, dt \\
    &\leq S(x(0), z(0))  + \int_{0}^{T} \xi(t) \, dt
    \end{aligned}
  \end{equation}
  holds for all $T > 0$,   $u\in \mathscr{U}$, $x(0) \in \mathcal{X}$ and $\bar{x} \in \mathcal{\bar{X}}$,  where $\xi(t)= \Xi(u, y, \bar{x})(t)$ and $x, z, y$ satisfy \eqref{eq: OLSystem} and \eqref{eq: aux}. In addition, it is called \emph{input (resp. output) strictly $\Xi$-dissipative} if \eqref{eq: strict_dissipation} holds without the
  term $\gamma_2(\cdot)$ (resp. $\gamma_1(\cdot)$). Strict $\Xi'$-dissipativity may be defined similarly as in Definition~\ref{def: dynamic_dissi} by requiring $\xi(t)= \Xi(u, y, x(0))(t)$.
\end{defn}

Definition~\ref{def: dynamic_strict_dissi} lists three different types of dissipativity strictness. It reminisces the definition of strict passivity in the literature; see, for example, \cite[Def.~6.3]{Kha02}. 
\begin{rem}\label{rem: ICD_example}
We have seen earlier in Example~\ref{exmp: ICD} that $\Sigma$ in \eqref{eq: numerical_exmp_ICD} is $\Xi'$-dissipative. We may now further characterise its dissipativity strictness via Definition~\ref{def: dynamic_strict_dissi}. Specifically, according to the first inequality in \eqref{eq: ICD_example}, $\Sigma$ is output strictly $\Xi'$-dissipative.
\end{rem}

Equipped with Definition~\ref{def: dynamic_strict_dissi} and Lemma~\ref{prop:dissi_stability}, we are ready to present the main result on
asymptotic stability of the closed-loop system $\Sigma_1 \| \Sigma_2$.

\begin{thm} \label{thm: diss} Suppose $\Sigma_1$ and $\Sigma_2$ in \eqref{eq: OLSystems} are zero-state detectable and there exists a causal operator $\Xi: \mathscr{U}_1 \times \mathscr{Y}_1 \times {\mathcal{X}_1} \to \boldsymbol{\rm L}_{1e}$ such that any one of the following dissipativity conditions holds:
  \begin{enumerate} \renewcommand{\theenumi}{\textup{(\roman{enumi})}}\renewcommand{\labelenumi}{\theenumi}
  \item \label{item: uy}
  $\Sigma_1$ is $\Xi'$-dissipative and $\Sigma_2$ is very strictly $(-\Xi\circ \Gamma)$-dissipative;

  \item \label{item: yu}
 $\Sigma_1$ is very strictly $\Xi'$-dissipative and $\Sigma_2$ is $(-\Xi\circ \Gamma)$-dissipative;

  \item \label{item: uu}
   $\Sigma_1$ is input strictly $\Xi'$-dissipative and $\Sigma_2$ is input strictly $(-\Xi\circ \Gamma)$-dissipative;
   
  \item \label{item: yy}
   $\Sigma_1$ is   output strictly $\Xi'$-dissipative and $\Sigma_2$ is output strictly $(-\Xi\circ \Gamma)$-dissipative.
  \end{enumerate}
   Furthermore, suppose the  corresponding storage functions $S_1$ and $S_2$ satisfy Assumption~\ref{assum:bounded_S} (resp. Assumption~\ref{assum:bounded_stable_S} and $(x^*, z^*) = (x_1^*, x_2^*, z_1^*, z_2^*) = (0, 0, 0, 0)$ is an equilibrium of $\Sigma_1 \| \Sigma_2$ 
   and auxiliary system \eqref{eq: two_aux} with $w_1 = 0$ and $w_2 = 0$).
Then, $x^* = (x_1^*, x_2^*) = (0, 0)$ is an asymptotically stable equilibrium  of the closed-loop
  system $\Sigma_1 \| \Sigma_2$ with $w_1 = 0$ and $w_2 = 0$.  Moreover, if Assumption~\ref{assum:bounded_S} (resp. Assumption~\ref{assum:bounded_stable_S})  holds for all $x \in \mathcal{X}$ and $z \in \mathcal{Z}$ for class $\mathcal{K}_{\infty}$ functions $\alpha$ and  $\beta$, then $x^* = 0$ is a globally asymptotically stable
  equilibrium.  
\end{thm}

\pf 
Let $S_1$ and $S_2$ satisfy Assumption~\ref{assum:bounded_S}. We show that $x^* = 0$ is an asymptotically stable equilibrium when \ref{item: uy} holds. Stability involving \ref{item: yu}, \ref{item: uu} or \ref{item: yy} may be established similarly. By hypothesis, we have
      \begin{align*}
    &  S_1(x_1(T), z_1(T)) \leq S_1(x_1(0), z_1(0))   \\
     & \hspace{3cm} + \int_{0}^{T} \Xi(u_1, y_1, {x_1(0)})(t) \, dt\\
       &     S_2(x_2(T), z_2(T))  \leq    S_2(x_2(0), z_2(0)) \\
        &-\int_{0}^{T}  \Xi(y_2, u_2, {x_1(0)})(t)\, dt -  \int_{0}^{T} [\gamma_1(u_2(t)) + \gamma_2(y_2(t))] \, dt
    \end{align*}
      for all $T > 0$, initial conditions $x_1(0)\in  \mathcal{X}_1$, $x_2(0)\in  \mathcal{X}_2$, and   input functions $u_1 \in  \mathscr{U}_1$, $u_2 \in \mathscr{U}_2$.
  By substituting the feedback equations
  \eqref{eq: FB}  with $w_1 = 0$ and $w_2 = 0$, i.e. $u_1 = y_2$, $u_2 = y_1$, into the conditions above and summing the inequalities,  we get
  \begin{multline}\label{eq: FB_diss}
  S_1(x_1(T), z_1(T)) + S_2(x_2(T), z_2(T))\\
    \begin{aligned}
      \leq &~ S_1(x_1(0), z_1(0)) + S_2(x_2(0), z_2(0))\\
      & -\int_{0}^{T} [\gamma_1(y_1(t)) +\gamma_2(y_2(t))] \, dt.
      \end{aligned}
  \end{multline}
  Let $S(x, z) = S_1(x_1, z_1) + S_2(x_2, z_2)$
  and
  \[
    \tilde{\xi}(t)= \tilde{\Xi}(w(t), y(t)) = -\gamma_1(y_1(t)) - \gamma_2(y_2(t)),
  \]
  where $x = (x_1, x_2)$, $z = (z_1, z_2)$, $w = (w_1, w_2)$, and $y = (y_1, y_2)$. Note that $\tilde{\Xi}(0, y(t)) \leq 0$,
  $\tilde{\Xi}(0, y(t)) = 0$ implies that $y(t) = 0$,  and $\tilde{\Xi}(0, y(t))$ is static and continuous in $y(t)\in \mathcal{Y}$  since $\gamma_1$ and $\gamma_2$ are  continuous  positive definite functions. Moreover, by \eqref{eq: FB_diss}, the closed-loop system $\Sigma_1 \| \Sigma_2$ with input $w = 0$
  and output $y$ is dissipative with a static supply rate $\tilde{\xi}(t)=\tilde{\Xi}(w(t), y(t))$ via the auxiliary system \eqref{eq: two_aux}, where the
  storage function $S$ is in $C_1$. Therefore, $x^* = 0$ is an asymptotically stable equilibrium via Lemma~\ref{prop:dissi_stability}. In addition, when Assumption~\ref{assum:bounded_S} holds for all $x_i \in \mathcal{X}_i$ and $z_i \in \mathcal{Z}_i$ and $\alpha_i, \beta_i$ are
  class $\mathcal{K}_\infty$ functions, global asymptotic stability of $x^*=0$ follows again from Lemma~\ref{prop:dissi_stability}. 

 Now suppose $S_1$ and $S_2$ satisfy Assumption~\ref{assum:bounded_stable_S} and $(x^*, z^*) = (0, 0)$ is an equilibrium. Using \eqref{eq: FB_diss}, one obtains that $(x^*, z^*)$ is Lyapunov stable by the standard Lyapunov theorem~\cite[Th. 3.2.4]{Sch17}. Let $S(x, z) = S_1(x_1, z_1) + S_2(x_2, z_2)$. Noting that \eqref{eq: FB_diss} and the zero-state detectability of $\Sigma_1$ and $\Sigma_2$ imply that the only solution $x$ that can stay identically in $\{x \in \mathcal{X} : \dot{S}(x, z) = 0\}$ is $x(t) = 0$ for all $t \ge 0$, asymptotic stability of $x^* = 0$ then follows from the LaSalle's invariance principle~\cite[Th. 4.4]{Kha02}.
  \hfill $\square$ \endpf

Notice that in conditions \ref{item: uy}-\ref{item: yy} of Theorem~\ref{thm: diss}, both $\Sigma_1$ and $\Sigma_2$ share complementary supply rates $\Xi'$ and $-\Xi \circ \Gamma$, which are used to characterise dynamic dissipativity.  In comparison with the Lyapunov stability result (Theorem~\ref{thm: Lya_sta}),
  Theorem~\ref{thm: diss} requires extra positive definite terms $\gamma_i(\cdot)$ as in \eqref{eq: strict_dissipation} for describing the strictness
  of dissipativity so that the stronger notion of asymptotic stability of $\Sigma_1 \| \Sigma_2$ can be established. It is worth noting that
  conditions \ref{item: uy}-\ref{item: yy} of Theorem~\ref{thm: diss} involve permutations of the positive definite terms $\gamma_i(\cdot)$ on the
  inputs and outputs of the open-loop systems. Some important special cases of Theorem~\ref{thm: diss} relating to the literature are detailed in
  Section~\ref{sec: examples}.

\begin{rem} \label{rem: diss}
  In the case where $\Sigma_2$ is static or stability of $x_2^* = 0$ is of no concern, the dissipativity conditions \ref{item: uy}-\ref{item: yy}  in Theorem~\ref{thm: diss}  for $\Sigma_2$ can be simplified by omitting $x_2$ as in \eqref{eq: static_dissi_ineq} and restricting $\mathcal{X}$ to be $\mathcal{X}_1$ in Assumption~\ref{assum:bounded_S} or \ref{assum:bounded_stable_S} and Theorem~\ref{thm: diss}. In this case, stability of $x_1^* = 0$ may be established with $S(x_1, z) = S_1(x_1, z_1) + S_2(z_2)$ by looking at the closed-loop map from $w_1$ to $y_1$. 
\end{rem}

Interestingly, asymptotic stability of the feedback system may be established using a type  of  strict dissipativity where the strictness is derived from
the state. In this case, the assumption on the zero-state detectability of $\Sigma_1$ and $\Sigma_2$ is not needed.
 
\begin{defn}[State-strict  dynamic dissipativity]
  Let $\Xi: \mathscr{U} \times \mathscr{Y} \times \Xbar \to \boldsymbol{\rm L}_{1e}$ be causal, and $\gamma$ be a  class $\mathcal{K}$ function. $\Sigma$ in \eqref{eq: OLSystem} is called \emph{state strictly $\Xi$-dissipative} if  there exist an auxiliary system~\eqref{eq: aux} and $S: \mathcal{X} \times \mathcal{Z} \to \Real$ such that
 \begin{equation}\label{eq: state-strict_dissipation}
  \begin{aligned}
  S(x(T), z(T)) &+ \int_{0}^{T} \gamma(\norm{x(t)}) \, dt \\
  & \leq S(x(0), z(0))  + \int_{0}^{T} \xi(t) \, dt
  \end{aligned}
\end{equation}
  holds for all $T > 0$, $u\in \mathscr{U}$, $x(0) \in \mathcal{X}$, and $\bar{x} \in \mathcal{\bar{X}}$, where $\xi(t)= \Xi(u, y, \bar{x})(t)$ and $x, z, y$ satisfy \eqref{eq: OLSystem} and \eqref{eq: aux}. Furthermore, $\Sigma$ is said to be \emph{state strictly $\Xi'$-dissipative} if \eqref{eq: state-strict_dissipation} holds for all $T > 0$, $u\in \mathscr{U}$, and $x(0) \in \mathcal{X}$, with $\xi(t)= \Xi(u, y, x(0))(t)$.
\end{defn}

\begin{thm}\label{thm: diss_state}
   The conclusions of Theorem~\ref{thm: diss} still hold if   the dissipativity conditions \ref{item: uy}-\ref{item: yy} are replaced by:
   \begin{enumerate}
    \setcounter{enumi}{4}
    \renewcommand{\theenumi}{\textup{(\roman{enumi})}}\renewcommand{\labelenumi}{\theenumi}
  \item    
  $\Sigma_1$ is state strictly  $\Xi'$-dissipative and $\Sigma_2$ is state strictly $(-\Xi\circ \Gamma)$-dissipative
   \end{enumerate}
  and the supposition of   zero-state detectability of $\Sigma_1$ and $\Sigma_2$ is removed.
\end{thm}
  \pf
  Using time-invariance of $\Sigma_1$ and $\Sigma_2$ and summing the two dissipation inequalities yield that
    \begin{multline*}
    S_1(x_1(t_2), z_1(t_2)) + S_2(x_2(t_2), z_2(t_2)) \\
    \begin{aligned}
    \leq &~ S_1(x_1(t_1), z_1(t_1)) + S_2(x_2(t_1), z_2(t_1)) \\
    &- \int_{t_1}^{t_2} (\gamma_1(\norm{x_1(t)}) + \gamma_2(\norm{x_2(t)})) \, dt
      \end{aligned}
  \end{multline*}
  along the solutions to  \eqref{eq: FB_state} and \eqref{eq: two_aux}. Dividing both sides by $t_2-t_1$ and taking  the limit as $t_2 \to t_1$ yields
  \begin{equation}\label{eq: FB_diss_Lyap}
    \begin{aligned}
      \frac{d}{dt} S(x, z)\leq -\gamma(\norm{x(t)})\leq 0
    \end{aligned}
  \end{equation} 
  along the solutions to \eqref{eq: FB_state} and \eqref{eq: two_aux}, where $S(x, z) = S_1(x_1, z_1) + S_2(x_2, z_2)$ and  $\gamma(\norm{x(t)})=\gamma_1(\norm{x_1(t)})+\gamma_2(\norm{x_2(t)})$. 
  
For the case where Assumption~\ref{assum:bounded_S} holds: By the same arguments as in the proof of
  Theorem~\ref{thm: Lya_sta}, $x^* = 0$ is an equilibrium of \eqref{eq: FB_state} with $w_1 = 0$ and $w_2=0$. Moreover, observe that $\gamma$ is a
  class $\mathcal{K}$ function because so are $\gamma_1$ and $\gamma_2$. It follows that $x^* = 0$ is an asymptotically stable equilibrium via
  \cite[Th.~4.1(iv)]{Haddad08} on noting \eqref{eq:partial_bounded_S} and \eqref{eq: FB_diss_Lyap}.

 Next, for the case when Assumption~\ref{assum:bounded_stable_S} holds and $(x^*, z^*) = (0, 0)$ is an equilibrium: Lyapunov stability of $(x^*, z^*) = (0, 0)$ follows from \eqref{eq: FB_diss_Lyap}. Since the only solution $x$ of $\Sigma_1 \| \Sigma_2$ that can stay identically in $\{x \in \mathcal{X} : \dot{S}(x, z) = 0\}$ is $x(t) = 0$ for all $t \ge 0$, asymptotic stability of $x^* = 0$ then holds by the LaSalle's invariance principle~\cite[Th. 4.4]{Kha02}.
  
  Finally, for the case when $\alpha$ and $\beta$ are class $\mathcal{K}_\infty$ functions, global asymptotic stability of
  $x^*=0$   can be concluded using similar arguments as above. \hfill $\square$
  \endpf
 
  \subsection{Exponential Stability}
  
While asymptotic stability guarantees convergence to the origin as time progresses, it gives no a priori rate of convergence. We may investigate  the
even stronger notion of exponential closed-loop stability  via dissipativity. To this end, a stronger notion of dissipativity called exponential dissipativity
is warranted.  In light of Definition~\ref{def: dynamic_dissi}, the exponential $\Xi$-dissipativity of a system  with  decay rate $\lambda>0$ can be naturally defined by
changing the dissipation inequality in \eqref{eq: dissipation} to  
\begin{equation}\label{eq: exponential_diss_ineq}
  \begin{aligned}
\hspace{-2mm} e^{\lambda T}S(x(T), z(T)) \leq S(x(0), z(0)) + \int_{0}^{T} e^{\lambda t}\xi(t) \, dt,
  \end{aligned}
\end{equation}
where $\xi(t)= \Xi(u, y, \bar{x})(t)$.  
Exponential dissipativity with respect to  static supply rates and dynamic supply rates  with a  quadratic form under $\bar{\mathcal{X}}=\emptyset$ has been  established in \cite{XHK05}. We note
that by requiring \eqref{eq: exponential_diss_ineq} for both $\Sigma_1$ and $\Sigma_2$, and under  mild assumptions on the storage functions $S_1$
and $S_2$, one can easily establish  exponential stability of $\Sigma_1 \| \Sigma_2$  via generic dynamic supply rates that need not have a quadratic form in a similar fashion to Theorem~\ref{thm: diss}. We omit the
details of such a result for brevity.
 
\section{Specialisation of the Main Results}\label{sec: examples}
 
In this section, we specialise the main results --- Theorems~\ref{thm: Lya_sta} and \ref{thm: diss} ---  from the preceding section to obtain several
corollaries pertinent to static and dynamic dissipativity results. 

\subsection{$(\Psi$, $\Pi$, $\Upsilon$, $\Omega)$-dissipativity}
 
The celebrated $(Q, \hat{S}, R)$-dissipativity \cite{HilMoy76} has made a profound impact on the theory of dissipativity over the past
half-century. It involves using  a static matrix triplet ($Q$, $\hat{S}$, $R$) in a static quadratic supply rate.  The following specialisation of Theorem~\ref{thm: diss} generalises the static matrix triplet ($Q$, $\hat{S}$, $R$) to a dynamic operator quadruplet $(\Psi$, $\Pi$, $\Upsilon$, $\Omega)$.

Let the operators $\Psi_i$, $\Pi_i$, $\Upsilon_i$, $\Omega_i$ be causal and time-invariant for  $i\in \{1, 2\}$ and let
\begin{align}\label{eq: dynamic_qsr}
  \Theta_i = \TwoTwo{ \Psi_i}{\Pi_i}{\Upsilon_i}{\Omega_i} 
\end{align}
The causal operators $\Theta_i$ for $i\in\{1,2\}$ do not need to be bounded on positive time support. Hence, they do not need to be stable operators. Furthermore, the operators $\Theta_i$ for $i\in\{1,2\}$ can be any causal system (whether static or dynamic, linear or non-linear) and do not need to have a state-space representation (e.g. may have pure derivatives).
Define the dynamic supply rates for $\Sigma_1$ and $\Sigma_2$ by
\begin{equation}\label{eq: dynamic_qsr_rate}
  \begin{aligned}
     \Xi_i (u_i, y_i, \xbar)    = \TwoOne{u_i}{y_i}^\top  \left(\Theta_i \TwoOne{u_i}{y_i}\right), 
  \end{aligned}
\end{equation}
 where $\Theta_i$ do not depend on $\bar{x}$. Such a special form of $\Xi$-dissipativity may be referred to as  ``$(\Psi, \Pi, \Upsilon, \Omega)$-dissipativity'' and the following   result can be derived.

\begin{thm}\label{thm: dynamic_qsr}
   Let $\Sigma_i$ in \eqref{eq: OLSystems} be zero-state detectable  and 
  $\Xi_i$-dissipative with $\Xi_i$ given by  \eqref{eq: dynamic_qsr_rate} and causal time-invariant $\Theta_i$ given by \eqref{eq: dynamic_qsr} for $i\in\left\{1, 2\right\}$.
  Let the corresponding storage functions $S_i$ satisfy Assumption~\ref{assum:bounded_S} (resp. Assumption~\ref{assum:bounded_stable_S} and suppose $(x^*, z^*) = (0, 0)$ is an equilibrium of \eqref{eq: FB_state} and \eqref{eq: two_aux} with $w_1=0$, $w_2=0$) and define $H = \STwoTwo{0}{I_p}{I_m}{0}$.  Suppose there exists $\tau > 0$ such that the causal operator
  \begin{equation}\label{eq: cor_dynamic_qsr}
    \Theta = - (\Theta_1 + \tau H^\top\Theta_2H)
  \end{equation}
  is  input strictly passive in the input-output sense\footnote[2]{ A causal operator $\Theta: \Ltwoe^{m+p} \to \Ltwoe^{m+p}$ is said to be \emph{input strictly passive in the input-output sense} if there exists $\delta>0$ such that $\langle f \,, \Theta f \rangle_T  \geq \delta \norm{f}_T^2 $ for all $f\in \Ltwoe^{m+p}$ and $T>0$.}.  Then $x^* = (x_1^*, x_2^*) = 0$ is an asymptotically stable equilibrium of the closed-loop system $\Sigma_1 \| \Sigma_2$ with $w_1 = 0$, $w_2 = 0$.   Moreover, if Assumption~\ref{assum:bounded_S} (resp. Assumption~\ref{assum:bounded_stable_S}) holds for all $x \in \mathcal{X}$ and
  $z \in \mathcal{Z}$ for class $\mathcal{K}_{\infty}$ functions $\alpha$ and  $\beta$,  $x^* = 0$ is a globally asymptotically stable
  equilibrium.
\end{thm}

\pf
  Since $\Theta_2 =  \frac{1}{\tau}H(-\Theta - \Theta_1)H^\top$, we have that $\Sigma_1$ and $\Sigma_2$ satisfy the following dissipation inequalities:
  \begin{align*}
    S_1(x_1(T), z_1(T)) &- S_1(x_1(0), z_1(0)) \\
    &\qquad \leq \int_{0}^{T} \TwoOne{u_1(t)}{y_1(t)}^\top  \left(\Theta_1 \TwoOne{u_1}{y_1}\right)(t)\,dt
  \end{align*}
and
 \begin{align*}
    &\tau S_2(x_2(T), z_2(T)) - \tau S_2(x_2(0), z_2(0)) \\
    &\qquad \leq -\int_{0}^{T} \TwoOne{y_2(t)}{u_2(t)}^\top  \left(\Theta \TwoOne{y_2}{u_2}\right)(t)\,dt \\
    &\qquad\quad -\int_{0}^{T} \TwoOne{y_2(t)}{u_2(t)}^\top  \left(\Theta_1 \TwoOne{y_2}{u_2}\right)(t)\,dt \\
    &\qquad \leq -\delta\norm{\TwoOne{y_2}{u_2}}^2_T 
                 -\int_{0}^{T} \TwoOne{y_2(t)}{u_2(t)}^\top  \left(\Theta_1 \TwoOne{y_2}{u_2}\right)(t)\,dt
  \end{align*}
  for some $\delta > 0$, where the last inequality follows from  input strict passivity of the operator $\Theta$. 
 Now let $\hat{S}_2=\tau S_2$. Since $\Xi_1(u_1, y_1, \xbar) = \STwoOne{u_1}{y_1}^\top  \left(\Theta_1 \STwoOne{u_1}{y_1}\right)$ 
  and $(-\Xi_1\circ\Gamma)(u_2, y_2, \xbar) = -\STwoOne{y_2}{u_2}^\top  \left(\Theta_1 \STwoOne{y_2}{u_2}\right)$ and both do not depend on $\xbar$,
  it can be seen that condition~\ref{item: uy} in Theorem~\ref{thm: diss} holds and the claim follows from the same theorem.  \hfill $\square$ \endpf

If the causal dynamic operators $\Psi_i$, $\Pi_i$, $\Upsilon_i$, $\Omega_i$ in \eqref{eq: dynamic_qsr} are LTI with frequency-domain representations
  $\Psi_i(\omega)$, $\Pi_i(\omega)$,  $\Upsilon_i(\omega)$, $\Omega_i(\omega)$ respectively, then $\Xi_i$-dissipativity in   Theorem~\ref{thm: dynamic_qsr} captures 
  the $(Q(\omega), S(\omega), R(\omega))$-dissipativity   notions in \cite{Griggs07, Patra11, Lanzon23} and also   relates to the theory of integral quadratic constraints
  (IQCs) \cite{MegRan97}. Such a connection is discussed in more detail in Section~\ref{sec: relation_IQCs}.

  While  $\Sigma_1$ and $\Sigma_2$ may be dissipative with respect to different supply rates in  Theorem~\ref{thm: dynamic_qsr}, the supply rates are related via the
   \emph{coupling term}~\eqref{eq: cor_dynamic_qsr} that requires $\Theta$ defined therein to be input strictly passive. Furthermore, if the  causal operators $\Psi_i$, $\Pi_i$,  $\Upsilon_i$, $\Omega_i$ in \eqref{eq: dynamic_qsr} are LTI with corresponding state-space realisations, the
  input strict passivity requirement on $\Theta$ can be easily tested  numerically via linear matrix inequalities (LMIs); see, for example, \cite[Sec.~3.1]{BLME07}.

  Several existing static dissipativity results in the literature may be established using  Theorem~\ref{thm: dynamic_qsr} with empty auxiliary
  systems, wherein $\mathcal{Z}_i=\emptyset$. First, the small-gain theorem in~\cite[Th.~8.2.1]{Sch17} assumes that $\Sigma_i$ is $\Xi_i$-dissipative with static ${\Theta}_i=\STwoTwo{r_i^2I}{0}{0}{-I}$  in
  \eqref{eq: dynamic_qsr_rate},  $r_i>0$ and
  $r_1r_2<1$.  Choosing  $\tau \in \left(r_1^2, {1}/{r_2^2}\right)$ in Theorem~\ref{thm: dynamic_qsr}  ensures that  coupling term \eqref{eq: cor_dynamic_qsr} is input strictly passive. The small-gain theorem in~\cite[Th.~8.2.1]{Sch17} is thus a specialisation of  Theorem~\ref{thm: dynamic_qsr}.

  Second,  the passivity theorem  in~\cite[Prop.~4.3.1(iv)]{Sch17}  uses a negative feedback configuration and assumes that $\Sigma_1$ and
  $-\Sigma_2$ are output strictly passive, i.e. $\Sigma_1$ is $\Xi_1$-dissipative and $-\Sigma_2$ is $\Xi_2$-dissipative with static $\Theta_i=\STwoTwo{0}{1/2I}{1/2I}{-\epsilon_iI}$ in \eqref{eq:
  dynamic_qsr_rate}  and
  $\epsilon_i>0$. Choosing $\tau=1$ in  Theorem~\ref{thm: dynamic_qsr}  ensures that coupling term
  \eqref{eq: cor_dynamic_qsr} is input strictly passive after absorbing the negative sign of $-\Sigma_2$ into $y_2$. Therefore, the passivity theorem in~\cite[Prop.~4.3.1(iv)]{Sch17}  is also a specialisation of Theorem~\ref{thm: dynamic_qsr}.

 Third, one can also derive the following immediate result on passivity indices. Let $\Sigma_1$ and $-\Sigma_2$ be input-feedforward and output-feedback
  passive, i.e.  $\Sigma_1$ is $\Xi_1$-dissipative and $-\Sigma_2$ is $\Xi_2$-dissipative   with static
  $\Theta_i=\STwoTwo{-\delta_i I}{1/2I}{1/2I}{-\epsilon_iI}$ in \eqref{eq: dynamic_qsr_rate}.  If $\delta_1 + \epsilon_2 > 0$ and $\delta_2 + \epsilon_1 > 0$, then   choosing $\tau=1$ ensures that \eqref{eq:
    cor_dynamic_qsr} is input strictly passive after absorbing the negative sign of $-\Sigma_2$ into $y_2$ and asymptotic stability of $\Sigma_1 \| (-\Sigma_2)$ can be established via Theorem~\ref{thm: dynamic_qsr}. Such a result reminisces the
  finite-gain input-output closed-loop stability result based on passivity indices in~\cite[Thm. 6.6.58]{Vid02}.

\subsection{Dissipation with Terminal Costs}

Specific types of dynamic supply rates have appeared in the study of finite-gain input-output stability of feedback systems via dissipation
inequalities~\cite{Sei15, SchVee18} as a means to recover the standard theory of IQCs~\cite{MegRan97}. In~\cite{Sei15, SchVee18}, feedback interconnections of a nonlinear system and an LTI system are considered and canonical
factorisations of the multipliers are crucial. Similar dynamic supply rates can also be located in~\cite{AMP16,Sch21}, where asymptotic stability of feedback systems is examined.

It is demonstrated in~\cite[Th.~30]{Sch21} that the dynamic dissipativity with terminal costs result \cite[Th.~13]{Sch21} can be used  to recover
the renowned IQC-based input-output stability result~\cite[Th.~1]{MegRan97} for a feedback interconnection of a stable nonlinearity and a stable
LTI system. Next, we demonstrate that \cite[Th.~13]{Sch21} on dissipativity with terminal
costs, restated in Corollary~\ref{cor: Scherer} below for convenience, is a specialisation of Theorem~\ref{thm: diss}.

\begin{cor} \label{cor: Scherer}
  Let $\Sigma_2$ be an LTI system with minimal realisation $(A, B, C, D)$. Let  an auxiliary system $\Phi$ be LTI with  realisation
  $\left(A_\Phi, \OneTwo{B_{\Phi_1}}{B_{\Phi_2}}, C_{\Phi}, \OneTwo{D_{\Phi_1}}{D_{\Phi_2}}\right)$ and state variable $z$ with $z(0)=0$. Given $P = P^\top$, suppose
  there exist $X = X^\top$, $Z = Z^\top$, and $\epsilon > 0$ such that $\Sigma_1$ satisfies  
  \begin{align*}
    z(T)^\top Z z(T) \leq  \int_0^T \left(\Phi\TwoOne{u_1}{y_1}\right)(t)^\top P \left(\Phi\TwoOne{u_1}{y_1}\right)(t) \, dt
  \end{align*}
  for all $T>0$, $u_1\in \mathscr{U}_1$, and $y_1$ being a solution to \eqref{eq: OLSystems},  
    and $\Sigma_2$ satisfies 
      \begin{multline*}
    \TwoOne{z(T)}{x_2(T)}^\top X \TwoOne{z(T)}{x_2(T)}  \leq  \TwoOne{0}{x_2(0)}^\top X \TwoOne{0}{x_2(0)} \\
    \begin{aligned}
    & - \int_0^T \left(\Phi\TwoOne{y_2}{u_2}\right)(t)^\top P
   \left(\Phi\TwoOne{y_2}{u_2}\right)(t) \, dt \\
   & - \epsilon \int_0^T (\norm{u_2(t)}^2 + \norm{y_2(t)}^2) \, dt
    \end{aligned}
  \end{multline*}
  for all $T>0$, $u_2\in \mathscr{U}_2$, $x_2(0)\in \mathcal{X}_2$, and $x_2$ and $y_2$ being a solution to \eqref{eq: OLSystems}.
  If $X + \diag(Z, 0) > 0$, then $x_2^* = 0$ is an asymptotically stable equilibrium of the closed-loop system $\Sigma_1 \| \Sigma_2$ with $w_1 = 0$
  and $w_2 = 0$.
\end{cor}

\pf
  Note that $\Sigma_2$ is zero-state detectable because $(A, B, C, D)$ is a minimal realisation. Let the supply rate
  \begin{align*}
   \Xi(u, y, \xbar)(t)  = \left(\Phi\TwoOne{u}{y}\right)(t)^\top P \left(\Phi\TwoOne{u}{y}\right)(t), 
  \end{align*}   
  where $\Xi$ does not depends on $\xbar$ and the storage functions
  \begin{align*}
    S_1(z)  = z^\top Z z~\text{and}~S_2(x_2, z)  = \TwoOne{z}{x_2}^\top X \TwoOne{z}{x_2}.
  \end{align*}
Here,  $\Sigma_1$ and $\Sigma_2$ ``share'' the same auxiliary system $\Phi$ and state $z$, and the   operator $\Xi$ is constructed from $\Phi$.  Since stability of $x_1^* = 0$ is of no concern, together with $X + \diag(Z, 0) > 0$, one can easily verify that Assumption~\ref{assum:bounded_stable_S}, 
  condition \ref{item: uy} in Theorem~\ref{thm: diss} and Remark~\ref{rem: diss}  hold, from which the result holds. \hfill $\square$
\endpf

\subsection{Dissipation for Affine-nonlinear Systems}

We show next that the dynamic dissipativity setting in \cite{XHK05}, where control-affine nonlinear systems are considered, may be recovered
from Theorems~\ref{thm: Lya_sta} and \ref{thm: diss}. To this end, consider   $\Sigma_1 \| \Sigma_2$, where $\Sigma_i$ is restricted
to be of the following affine forms:
\begin{align}\label{eq: affine_OLsystem}
  \Sigma_i :\;
  \begin{array}{l}
    \dot{x}_i(t) = {F_i(x_i(t))} + G_i(x_i(t))u_i(t), \\ 
    y_i(t) = { H_i(x_i(t))} + J_i(x_i(t))u_i(t)  
  \end{array}
\end{align}
and   $F_i(\cdot)$, $G_i(\cdot)$, $H_i(\cdot)$ and $J_i(\cdot)$ map $\mathcal{X}_i$ to real matrices and vectors with compatible dimensions. Next, associate $\Sigma_1$ and
$\Sigma_2$ with an auxiliary system $\Phi = \STwoOne{u_\Phi}{y_\Phi} \mapsto {\phi}$ represented by the following affine form:
\begin{align}\label{eq: auxiliary_common}
 \hspace{-3.5mm}  
  \begin{array}{l}
  \dot{z}(t) = {F_\Phi(z(t))} + G_\Phi(z(t))u_\Phi(t) + I_\Phi(z(t))y_\Phi(t)\\ 
 {\phi}(t) = {H_\Phi(z(t))} + J_\Phi(z(t))u_\Phi(t)+ K_\Phi(z(t))y_\Phi(t)  
\end{array}
\end{align}
with $z(0)=0$, and $F_\Phi(\cdot), G_\Phi(\cdot), I_\Phi(\cdot), H_\Phi(\cdot), J_\Phi(\cdot)$ and $K_\Phi(\cdot)$ map $\mathcal{Z}$ to real matrices and vectors with compatible
dimensions. In \cite{XHK05}, the same auxiliary system $\Phi$ is adopted for both $\Sigma_1$ and $\Sigma_2$, with inputs
$\STwoOne{u_\Phi}{y_\Phi}=\STwoOne{u_1}{y_1}$ and $\STwoOne{u_\Phi}{y_\Phi}=\STwoOne{y_2}{u_2}$, respectively.  Define the following two operators: 
\begin{equation}\label{eq: nonlinear_qsr}
  \begin{aligned}
  \Xi_1 (u_1, y_1, \xbar)(t) & = \left(\Phi\STwoOne{u_1}{y_1}\right)(t)^\top P_1\left(\Phi\STwoOne{u_1}{y_1}\right)(t),\\
   \Xi_2 (u_2, y_2, \xbar)(t) & =    \left(\Phi\STwoOne{y_2}{u_2}\right)(t)^\top P_2\left(\Phi\STwoOne{y_2}{u_2}\right)(t),
  \end{aligned} 
\end{equation}
where $P_1=P_1^\top$ and $P_2=P_2^\top$ are static matrices and both $\Xi_1$ and $\Xi_2$ are independent of $\xbar$. The following corollary, which first appeared in \cite[Th.~3.2]{XHK05},   is a specialisation of Theorem~\ref{thm: Lya_sta}.

\begin{cor}\label{cor: quadratic_dynamic_diss}
  Suppose there exist a common auxiliary system $\Phi$ in (\ref{eq: auxiliary_common}), $P_1=P_1^\top$ and $P_2=P_2^\top$ such that $\Sigma_1$ is
  $\Xi_1$-dissipative and $\Sigma_2$ is $\Xi_2$-dissipative with quadratic dynamic supply rates of the form \eqref{eq: nonlinear_qsr}. Suppose further that the corresponding storage functions $S_1$ and
  $S_2$ satisfy Assumption~\ref{assum:bounded_S}. If there exists $\tau > 0$ such that $P_1 + \tau P_2\leq 0$, then $x^* = (x_1^*, x_2^*) = 0$ is a
  Lyapunov stable equilibrium of $\Sigma_1 \| \Sigma_2$ with $w_1 = 0$
  and $w_2 = 0$.
\end{cor}
\pf
Let $P = P_1 + \tau P_2$ and note that  $P_2 = (P - P_1)/\tau$. By hypothesis, $\Sigma_1$ and $\Sigma_2$ satisfy the following dissipation inequalities
  \begin{equation*}
    \begin{aligned}
      S_1(x_1(T), z_1&(T)) \leq   S_1(x_1(0), 0) +   \int_{0}^{T} \Xi_1(u_1, y_1, \xbar)(t) \, dt,\\
     \tau S_2(x_2(T), z_2&(T)) \leq   \tau S_2(x_2(0), 0) \\
    &- \int_{0}^{T} (\Phi\STwoOne{y_2}{u_2} \vphantom{\Phi\TwoOne{y_2}{u_2}})(t)^\top (P_1-P)\left(\Phi\STwoOne{y_2}{u_2}\right)(t) \, dt \\
                  \leq &~\tau S_2(x_2(0), 0) - \int_{0}^{T}  
                   \Xi_1\circ \Gamma(u_2, y_2, \xbar)(t) \, dt,                
    \end{aligned}
  \end{equation*}
  where the last inequality follows from the fact $P\leq 0$. By defining the supply rate $\Xi(u, y,\xbar)(t) =\Xi_1(u, y,\xbar)(t)$, we conclude that $x^* = 0$ is
  a Lyapunov stable equilibrium with $w_1 = 0$, $w_2 = 0$ via an invocation Theorem~\ref{thm: Lya_sta}.\hfill $\square$ \endpf

  The asymptotic stability version of Corollary~\ref{cor: quadratic_dynamic_diss} can be  similarly established by  specialising Theorem~\ref{thm: diss}. It is
  omitted here for  conciseness.

The notion of dynamic dissipativity introduced in this paper is more general than that in \cite{XHK05} as explained next. First, the supply rate \eqref{eq: nonlinear_qsr}
considered in \cite{XHK05} is constructed directly from the auxiliary system  $\Phi$, while we treat the supply rate $\Xi$ and auxiliary system $\Phi$ as two independent objects,
cf. Fig.~\ref{fig: aux}. Second, the same auxiliary system $\Phi$ in \eqref{eq: auxiliary_common} is adopted for both $\Sigma_1$ and $\Sigma_2$ for feedback
stability analysis in \cite{XHK05}, while auxiliary systems  $\Phi_i$ in \eqref{eq: two_aux} can be different for $\Sigma_1$ and $\Sigma_2$ in Theorems~\ref{thm: Lya_sta} and
\ref{thm: diss}. Third, dissipativity of the quadratic form in \eqref{eq: nonlinear_qsr} is considered in \cite{XHK05} and defined  via control-affine auxiliary
systems in \eqref{eq: auxiliary_common}, whereas in this paper, dissipativity of the general form in (\ref{eq: dissipation}) via general auxiliary
systems in (\ref{eq: aux}) are considered  for general nonlinear systems in \eqref{eq: OLSystem}.

\section{Relations with Integral Quadratic Constraint Theory}\label{sec: relation_IQCs}

This section elaborates the relation between the dissipativity results for robust feedback Lyapunov-type stability in Section~\ref{sec: FB} and IQC
results for robust feedback finite-gain input-output stability~\cite{MegRan97, Kho22}.

We first  define some notation required for finite-gain input-output feedback stability. Denote by $\Ltwo^n$ the set of $\Real^n$-valued Lebesgue square-integrable functions:
\begin{align*}
  \Ltwo^n = \Big\{v: \Real  \to \Real^n\ \mid \ \|v\|^2_{\boldsymbol{\rm L}_2} = \int_{-\infty}^\infty \norm{v(t)}^2 \, dt< \infty  \Big\}.
\end{align*}
Let $\Ltwop^n = \{v \in \Ltwo^n : v(t) = 0 \text{ for all } t < 0\}$.  Recall the truncation operator $P_T$ 
and define the extended space
\begin{align*}
 \Ltwoe^n = \{v : \Real \to \Real^n \ |\  P_Tv \in \Ltwop^n \; \forall T \in [0, \infty)\}.
\end{align*}
 Given $v, w \in \Ltwoe^n$, let
$\langle v,  w \rangle_T = \int_0^T v(t)^\top w(t) \, dt$ and $\|v\|_T^2 = \langle v,  v \rangle_T$.
An operator $\Sigma : \Ltwoe^n \to \Ltwoe^n$ is said to be \emph{incrementally $\Ltwoe$-bounded} if
\[
\sup_{\substack{ T > 0; P_Tx \neq P_T y; \\ x, y \in \Ltwoe^n}} \frac{\|P_T (\Sigma x - \Sigma y) \|_{\boldsymbol{\rm L}_2}}{\|P_T(x - y)\|_{\boldsymbol{\rm L}_2}} < \infty.
\]
Note that an incrementally $\Ltwoe$-bounded
$\Sigma$ is necessarily causal~\cite[Prop.~2.1.6]{Sch17}. A causal $\Sigma$ is called \emph{bounded} if its bound~\cite[Sec.~2.4]{Wil71} is
finite, i.e.
\begin{align*}  
\|\Sigma\| = \sup_{\substack{T > 0; \|u\|_T \neq 0;\\ u \in \Ltwoe^n}} \frac{\|\Sigma u\|_T}{\|u\|_T} = \sup_{0 \neq u \in \Ltwop^n} \frac{\|\Sigma u\|_{\boldsymbol{\rm L}_2}}{\|u\|_{\boldsymbol{\rm L}_2}} < \infty.
\end{align*}

The following definitions of well-posedness and finite-gain input-output stability for feedback system $\Sigma_1 \| \Sigma_2$ are standard.

\begin{defn} \label{def: FB} $\Sigma_1 \| \Sigma_2$ is said to be \emph{well-posed} if the map $(u_1, u_2) \mapsto (w_1, w_2)$ defined by
  \eqref{eq: FB} has a causal inverse on $\Ltwoe^{m+p}$. $\Sigma_1 \| \Sigma_2$ is said to be finite-gain $\Ltwop$-\emph{stable} if it is well-posed
  and the map
$\STwoOne{w_1}{w_2} \in \Ltwoe^{m+p} \mapsto  \STwoOne{u_1}{u_2}   \in \Ltwoe^{m+p}$ is bounded, i.e. there exists
  $C > 0$ such that
\begin{align*}  
\begin{split}
\int_0^T \norm{u_1(t)}^2 + \norm{u_2(t)}^2 \, dt 
\leq C \int_0^T \norm{w_1(t)}^2 + \norm{w_2(t)}^2 \, dt
\end{split}
\end{align*}
for all  $w_1 \in  \Ltwoe^m,  w_2 \in \Ltwoe^p$ and $T > 0$.
\end{defn}

The above notions of feedback well-posedness and finite-gain input-output stability are well studied; see~\cite{Wil71, DesVid75,
  GeoSmi97}. Define the  extended graph of $\Sigma_1$ as $\graphe{\Sigma_1} = \left\{\STwoOne{u_1}{y_1} \in \Ltwoe^{m+p} : y_1 = \Sigma_1 u_1 \right\}$.  Likewise, define the extended inverse
graph of $\Sigma_2$ as $\graphei{\Sigma_2} = \left\{\STwoOne{y_2}{u_2} \in \Ltwoe^{m+p} : y_2 = \Sigma_2 u_2 \right\}$.

We restate below an IQC result from \cite{Kho22} for comparison purposes.

\begin{thm}[\hspace{1sp}{\cite[Th.~III.1]{Kho22}}] \label{thm: hardIQC}
  Given causal systems $\Sigma_1: \Ltwoe^m \to \Ltwoe^p$ and $\Sigma_2: \Ltwoe^p \to \Ltwoe^m$ satisfying
  $\Sigma_i 0 = 0$, $i \in \left\{1, 2\right\}$, suppose $\Sigma_1 \| \Sigma_2$ is well-posed and there exist incrementally bounded multipliers
  $\Psi : \Ltwoe^{m + p} \to \Ltwoe^q$ and $\Pi : \Ltwoe^{m + p} \to \Ltwoe^q$ such that
  \begin{equation} \label{eq: graph_sep}
    \begin{aligned}
  \langle \Psi v_1, \Pi v_1 \rangle_T & \geq 0 \; \forall v_1 \in \graphe{\Sigma_1}, T > 0 \\
    \quad \text{and} \quad \langle \Psi v_2, \Pi v_2 \rangle_T & \leq -\epsilon \|v_2\|_T^2
    \; \forall v_2 \in \graphei{\Sigma_2}, T > 0.
    \end{aligned}
  \end{equation}
  Then $\Sigma_1 \| \Sigma_2$ is finite-gain $\Ltwop$-stable.
\end{thm}

Observe that by taking a quadratic supply rate of the form in \eqref{eq: quadratic_supply}, Theorems~\ref{thm: diss} and~\ref{thm: hardIQC} are
similar on important grounds and differ in a few significant aspects. In terms of similarities, Theorem~\ref{thm: hardIQC} relies on quadratic graph
separation  enforced by \eqref{eq: graph_sep}. Such a separation is captured by  conditions~\ref{item: uy}-\ref{item: yy} in Theorem~\ref{thm: diss} with the aid of storage
functions that possess properties listed therein. On the other hand, some important differences include:
\begin{enumerate} \renewcommand{\theenumi}{\textup{(\roman{enumi})}}\renewcommand{\labelenumi}{\theenumi}
\item The IQC based Theorem~\ref{thm: hardIQC} makes use of a \emph{quadratic} form involving incrementally bounded multipliers, while the supply rate in the
  dissipativity based Theorem~\ref{thm: diss} may accommodate more general forms;

\item Theorem~\ref{thm: hardIQC} is an input-output feedback stability result, whereas Theorem~\ref{thm: diss} is a Lyapunov-type stability result on the state
  of the closed-loop system;

\item Theorem~\ref{thm: diss} requires the existence of storage functions that satisfy several properties. Therefore, in practice, the conditions in
  Theorem~\ref{thm: hardIQC} may be easier to verify. In particular, the use of dynamic multipliers is both natural and well known in the theory of
  IQCs as a means to reduce conservatism~\cite{MegRan97, Kho22}. By contrast, introducing dynamics into the supply rate in a dissipation inequality
  often complicates the search for a suitable storage function.
   Fortunately, the distinct auxiliary systems aid with the satisfaction of the dissipation inequalities. 
  
  Theorem~\ref{thm: hardIQC} is a \emph{hard} (a.k.a. unconditional) IQC
  theorem, where the integrals are taken from $0$ to $T$ for all $T > 0$ for signals in extended spaces~\cite{MJKR11}. This has been shown
  in~\cite{Kho22} to be recoverable by a more powerful \emph{soft} (a.k.a. conditional) IQC theorem~\cite[Th.~IV.2]{Kho22}, where integrals are
  taken from $0$ to $\infty$ for square-integrable signals, when equipped with homotopies that are continuous in a gap distance
  measure~\cite{GeoSmi97}. There, the use of noncausal multipliers is readily accommodated.
\end{enumerate}

Despite the aforementioned dissimilarities, Corollary~\ref{cor: Scherer}, or its sister result~\cite[Th.~13]{Sch21}, as a specialised form of
Theorem~\ref{thm: diss}, has been shown in~\cite[Th.~30]{Sch21} to recover a limited version of the soft IQC theorem for a feedback
interconnection of a stable nonlinearity and a stable LTI system~\cite[Th.~1]{MegRan97}. The proof relies on the fact that dissipativity of the
stable LTI component with a quadratic storage function is equivalent to one that involves the additional exogenous signals $w_1$ and $w_2$ in
Fig.~\ref{fig: feedback}; see~\cite[Lem.~17]{Sch21} and~\cite[Lem.~1]{Sei15}. Such a result is not known to hold for nonlinear systems, and hence
in the nonlinear setting, dissipativity and IQC approaches to feedback stability analysis remain distinct.

It is worth noting that certain types of dissipation inequalities may be used to show input-output (finite-gain) stability as mentioned above; see, e.g., \cite{Sei15} and \cite{SchVee18}. Since such results typically involve some LTI dynamics and input-output stability is not the main
focus of this paper, we do not discuss them here in detail. It is known that under certain conditions, global exponential stability implies
input-output finite-gain stability~\cite[Sec.~6.3]{Vid02}\cite[Sec.~7.6]{Haddad08}. On the other hand, for Lur'e feedback systems involving a
static nonlinearity, it has been shown that under certain Lipschitz continuity (resp. boundedness) condition, input-output finite gain feedback
stability implies global attractivity~\cite[Sec.~6.3]{Vid02} (resp. exponential stability~\cite[Prop.~1]{MegRan97}) of the origin.
 
\section{A Numerical Example}\label{sec: simu}
This section  provides a numerical example to demonstrate  the stability result established in Theorem~\ref{thm: diss}.

Consider a feedback system 
$\Sigma_1 \| \Sigma_2$ in  Fig.~\ref{fig: feedback}, where $\Sigma_1=u_1\mapsto y_1$ is provided  in  \eqref{eq: numerical_exmp_ICD} with $x(0) = \STwoOne{x_1(0)}{x_2(0)}\in \mathcal{X}= \Real^2$ and $\Sigma_2=u_2\mapsto y_2$ is described by
\begin{equation*} 
  \Sigma_2:\;
  \begin{aligned}
  \dot{x}_{3}(t) &= - 5x_{3}(t) -\psi_2(x_3(t))   +u_2(t)\\
y_2(t)&= x_3(t)- 0.2 u_{2}(t)
\end{aligned}
\end{equation*}
with $x_3(0)\in \Real$, where $\psi_2: \mathbb{R}\rightarrow\mathbb{R}$ is locally  integrable and satisfies $\psi_2(0)=0$ and ${\psi}_2(r) r \geq 0$ for all $r\in\mathbb{R}$. We have shown in Example~\ref{exmp: ICD} and Remark~\ref{rem: ICD_example} that $\Sigma_1$ is  output strictly $\Xi'$-dissipative with respect to the supply rate $\Xi(u_1, y_1, x(0))$ provided in \eqref{eq: exmple_ICD_supply}. Next, we attempt to validate a complementary dissipation inequality for $\Sigma_2$. Consider the candidate storage function $S_2(x_3, z_2)= \frac{1}{2}x_3^{2}+\frac{1}{2}z_2^2$. The complementary supply rate  $(-\Xi\circ \Gamma)=\SThreeOne{u_2}{y_2}{\xbar}\to \xi_2$ is given by 
\begin{equation*}
  -\Xi\circ \Gamma:\;
  \begin{aligned}
\dot{z}_2(t) &= -  z_2(t)  +  y_2(t), \quad z_2(0) = \SOneTwo{0}{1} \bar{x} \\ 
\xi_2(t)&=  -y_2(t)(3z_2(t)+u_2(t))
  \end{aligned}
\end{equation*}
with $\bar{x} \in \Real^2$.
Observe that for all $x_3(0) \in \Real$ and $\bar{x} \in \Real^2$, 
\begin{align*}
    &\frac{d}{dt}S_2(x_3, z_2) + y_2^2 -\xi_2 = x_3\dot{x}_3 + z_2\dot{z}_2+     y_2^2 -\xi_2\\
    =&-5x_3^2 -x_3\psi_2(x_3) + u_2x_3 -z_2^2 + z_2y_2 +      y_2^2 +y_2(3z_2+u_2)\\
     \leq &   -  5x_3^2 + u_2\left(y_2+0.2 u_2\right) -z_2^2  +y_2^2 + 4z_2y_2 + u_2y_2 \\
    =&    -z_2^2 +  4z_2y_2  +y_2^2 + 0.2 u_2^2  + 2u_2y_2 - 5\left(y_2+0.2 u_2\right)^2\\
     =&    -z_2^2   + 4z_2y_2 - 4y_2^2 =    -(z_2-2y_2)^2\leq 0
\end{align*}
along the solutions to $\Sigma_2$ and $-\Xi\circ \Gamma$. This indicates that $\Sigma_2$ is output strictly $(-\Xi\circ \Gamma)$-dissipative by Lemma~\ref{prop: differential_dissi}. Applying Theorem~\ref{thm: diss}\ref{item: yy}, it may then be concluded that the equilibrium $(x_1^*, x_2^*, x_3^*) = (0, 0, 0)$ of $\Sigma_1 \| \Sigma_2$ is globally asymptotically stable.

\begin{figure}[htb] 
    \vspace{-2mm}
\begin{minipage}[b]{\linewidth}
  \hspace{-0.9cm}\includegraphics[width=1.2\textwidth]{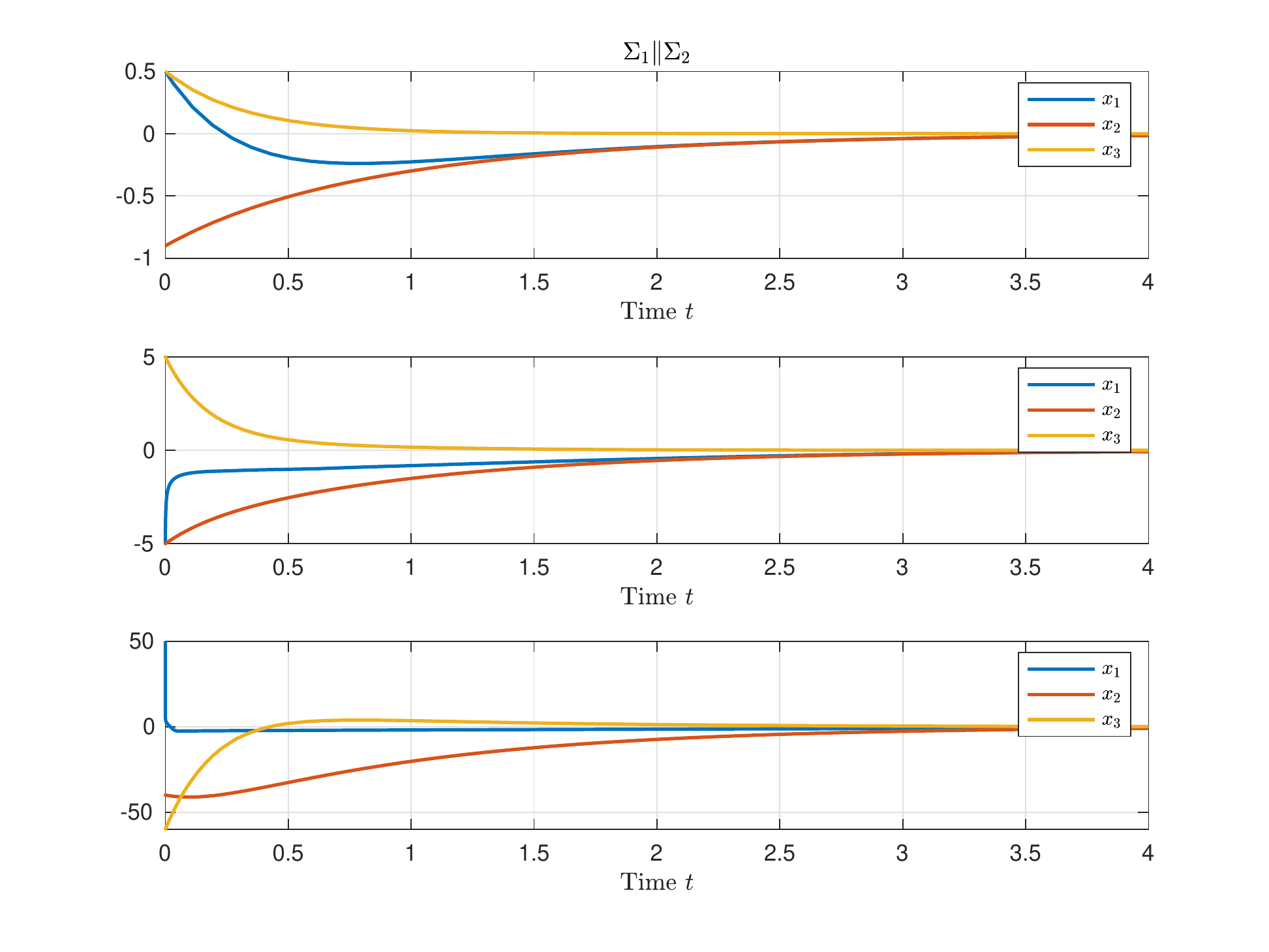}
 \end{minipage}
    \vspace{-8mm}
    \caption{Simulation results of the state trajectories of $\Sigma_1 \| {\Sigma}_2$.}\label{Fig: Sigma_2}
\end{figure}

As a simulation example,  for $\Sigma_1$, let $a=1$, $N=0$, $M=2$, $b_{k}=1$ for all $k$ and suppose $\psi$ is the saturation function described by $\psi(r)=  \min(\max(r, -5), 5)$. For $\Sigma_2$, suppose $\psi_2(r)=  \min(\max(r, -8), 8)$. The state trajectories  of $\Sigma_1 \| \Sigma_2$  are illustrated by Fig.~\ref{Fig: Sigma_2} under three different sets of initial conditions.
 
\section{Conclusion}\label{sec: conclusion}

In this paper, a general notion of dissipativity with dynamic supply rates was introduced for nonlinear systems, extending the notion of classical dissipativity. Lyapunov and asymptotic stability analyses were performed for feedback interconnections of two
dissipative systems satisfying dissipativity with respect to dynamic supply rates. In these results, both dynamical systems are characterised by compatible dissipation inequalities with respect to ``coupled''
dynamic supply rates.  Satisfaction of the dissipation inequalities is aided by the dynamics of possibly distinct auxiliary systems. The results were shown to recover several knowns results in the
literature. A noteworthy specialisation of the results is a  simple coupling test to verify whether the feedback interconnection of two nonlinear systems, each satisfying independent  $(\Psi, \Pi, \Upsilon, \Omega)$-dissipation inequalities, is   asymptotically stable. This coupling test is simple to compute if the supply rate operators are chosen to be LTI.  Moreover, a meaningful comparison with the integral quadratic constraint based input-output approach to feedback stability was
made. 

Future research directions include exploring physically illustrating examples for specific dynamic supply rates, such as those manifesting
negative imaginary dynamics, and developing dissipativity with dynamic supply rates for more general systems, for example those taking hybrid forms and large-scale interconnected networks.

\bibliographystyle{apalike}

\begin{thebibliography}{}

\bibitem[Angeli, 2006]{Angeli06}
Angeli, D. (2006).
\newblock Systems with counterclockwise input-output dynamics.
\newblock {\em IEEE Trans. Autom. Contr.}, 51(7):1130--1143.

\bibitem[Arcak et~al., 2016]{AMP16}
Arcak, M., Meissen, C., and Packard, A. (2016).
\newblock {\em Networks of Dissipative Systems: Compositional Certification of
  Stability, Performance, and Safety}.
\newblock Springer International Publishing AG, Cham, Switzerland.

\bibitem[Bao and Lee, 2007]{BaoLee07}
Bao, J. and Lee, P. (2007).
\newblock {\em Process Control}.
\newblock Springer-Verlag.

\bibitem[Bhowmick and Lanzon, 2024]{Bhowmick24}
Bhowmick, P. and Lanzon, A. (2024).
\newblock Dynamic dissipative characterisation of time-domain input-output
  negative imaginary systems.
\newblock {\em Automatica}, 164:111620 (1--14).

\bibitem[Boyd et~al., 1994]{BGFB94}
Boyd, S., Ghaoui, L.~E., Feron, E., and Balakrishan, V. (1994).
\newblock {\em Linear Matrix Inequalities in System and Control Theory}.
\newblock SIAM.

\bibitem[Brogliato et~al., 2007]{BLME07}
Brogliato, B., Lozano, R., Maschke, B., and Egeland, O. (2007).
\newblock {\em Dissipative systems analysis and control}.
\newblock Springer.

\bibitem[Cantoni et~al., 2013]{CJKh13}
Cantoni, M., J\"{o}nsson, U.~T., and Khong, S.~Z. (2013).
\newblock Robust stability analysis for feedback interconnections of
  time-varying linear systems.
\newblock {\em SIAM J. Control Optim.}, 51(1):353--379.

\bibitem[Carrasco and Seiler, 2019]{CarSei19}
Carrasco, J. and Seiler, P. (2019).
\newblock Conditions for the equivalence between {IQC} and graph separation
  stability results.
\newblock {\em Int. J. Control}, 92(12):2899--2906.

\bibitem[Chellaboina et~al., 2005]{XHK05}
Chellaboina, V., Haddad, W.~M., and Kamath, A. (2005).
\newblock Dynamic dissipativity theory for stability of nonlinear feedback
  dynamical systems.
\newblock In {\em Proc. 44th IEEE Conf. Decision and Control}, pages
  4748--4753, Seville, Spain.

\bibitem[Chen et~al., 2021]{Chen:20}
Chen, C., Zhao, D., Chen, W., Khong, S.~Z., and Qiu, L. (2021).
\newblock Phase of nonlinear systems.
\newblock {\em arXiv preprint arXiv:2012.00692}.

\bibitem[Crouch and van~der Schaft, 1987]{Crouch:87}
Crouch, P.~E. and van~der Schaft, A.~J. (1987).
\newblock {\em Variational and {H}amiltonian control systems}.
\newblock Springer-Verlag, Heidelberg, Germany.

\bibitem[Desoer and Vidyasagar, 1975]{DesVid75}
Desoer, C.~A. and Vidyasagar, M. (1975).
\newblock {\em Feedback Systems: Input-Output Properties.}
\newblock Academic Press, New York.

\bibitem[Doyle et~al., 1993]{DGS93}
Doyle, J.~C., Georgiou, T.~T., and Smith, M.~C. (1993).
\newblock The parallel projection of operators of a nonlinear feedback system.
\newblock {\em Syst. Control Lett.}, 20:79--85.

\bibitem[Forni and Sepulchre, 2013]{Forni:13}
Forni, F. and Sepulchre, R. (2013).
\newblock On differentially dissipative dynamical systems.
\newblock {\em IFAC Proceedings Volumes}, 46(23):15--20.

\bibitem[Forni and Sepulchre, 2018]{Forni18}
Forni, F. and Sepulchre, R. (2018).
\newblock Differential dissipativity theory for dominance analysis.
\newblock {\em IEEE Trans. Autom. Contr.}, 64(6):2340--2351.

\bibitem[Georgiou and Smith, 1997]{GeoSmi97}
Georgiou, T.~T. and Smith, M.~C. (1997).
\newblock Robustness analysis of nonlinear feedback systems: {An} input output
  approach.
\newblock {\em IEEE Trans. Autom. Contr.}, 42:1200--1221.

\bibitem[Ghallab and Petersen, 2022]{Ghallab22}
Ghallab, A.~G. and Petersen, I.~R. (2022).
\newblock Negative imaginary systems theory for nonlinear systems: a
  dissipativity approach.
\newblock {\em arXiv preprint, arXiv:2201.00144}.

\bibitem[Green and Limebeer, 1995]{GreLim95}
Green, M. and Limebeer, D. J.~N. (1995).
\newblock {\em Linear Robust Control}.
\newblock Information and System Sciences. Prentice-Hall.

\bibitem[Griggs et~al., 2007]{Griggs07}
Griggs, W.~M., Anderson, B. D.~O., and Lanzon, A. (2007).
\newblock A ``mixed'' small gain and passivity theorem in the frequency domain.
\newblock {\em Syst. Control Lett.}, 56(9-10):596--602.

\bibitem[Griggs et~al., 2009]{Griggs09}
Griggs, W.~M., Anderson, B. D.~O., Lanzon, A., and Rotkowitz, M. (2009).
\newblock Interconnections of nonlinear systems with ``mixed'' small gain and
  passivity properties and associated input-output stability results.
\newblock {\em Syst. Control Lett.}, 58(4):289--295.

\bibitem[Haddad and Chellaboina, 2008]{Haddad08}
Haddad, W.~M. and Chellaboina, V. (2008).
\newblock {\em Nonlinear Dynamical Systems and Control: {A} {L}yapunov-based
  approach}.
\newblock Princeton University Press.

\bibitem[Hilborne and Lanzon, 2022]{Hilborne22}
Hilborne, P. and Lanzon, A. (2022).
\newblock On local input-output stability of nonlinear feedback systems via
  local graph separation.
\newblock {\em {IEEE} Control Systems Letters}, 6:2894--2899.

\bibitem[Hill and Moylan, 1976]{HilMoy76}
Hill, D. and Moylan, P. (1976).
\newblock The stability of nonlinear dissipative systems.
\newblock {\em IEEE Trans. Autom. Contr.}, 21(5):708--711.

\bibitem[Hill and Liu, 2022]{Hill22}
Hill, D.~J. and Liu, T. (2022).
\newblock Dissipativity, stability, and connections: {P}rogress in complexity.
\newblock {\em IEEE Control Systems Magazine}, 42(2):88--106.

\bibitem[Hill and Moylan, 1980]{HilMoy80}
Hill, D.~J. and Moylan, P.~J. (1980).
\newblock Dissipative dynamical systems: Basic input-output and state
  properties.
\newblock {\em Journal of the Franklin Institute}, 309(5):327--357.

\bibitem[Iwasaki and Hara, 1998]{IwaHar98}
Iwasaki, T. and Hara, S. (1998).
\newblock Well-posedness of feedback systems: Insights into exact robustness
  analysis and approximate computations.
\newblock {\em IEEE Trans. Autom. Contr.}, 43(5):619--630.

\bibitem[Iwasaki and Hara, 2005]{IwaHar05}
Iwasaki, T. and Hara, S. (2005).
\newblock Generalized {KYP} lemma: {U}nified frequency domain inequalities with
  design applications.
\newblock {\em IEEE Trans. Autom. Contr.}, 50(1):41 -- 59.

\bibitem[Khalil, 2002]{Kha02}
Khalil, H.~K. (2002).
\newblock {\em Nonlinear Systems}.
\newblock Prentice Hall, 3rd edition.

\bibitem[Khong, 2022]{Kho22}
Khong, S.~Z. (2022).
\newblock On integral quadratic constraints.
\newblock {\em IEEE Trans. Autom. Contr.}, 67(3):1603--1608.

\bibitem[Khong and Kao, 2021]{KhoKao21}
Khong, S.~Z. and Kao, C.-Y. (2021).
\newblock Converse theorems for integral quadratic constraints.
\newblock {\em IEEE Trans. Autom. Contr.}, 66(8):3695--3701.

\bibitem[Khong and Kao, 2022]{KhoKao22}
Khong, S.~Z. and Kao, C.-Y. (2022).
\newblock Addendum to ``{Converse} theorems for integral quadratic
  constraints''.
\newblock {\em IEEE Trans. Autom. Contr.}, 67(1):539--540.

\bibitem[Khong and Lanzon, 2024]{KhongLanzon24}
Khong, S.~Z. and Lanzon, A. (2024).
\newblock Connections between integral quadratic constraints and dissipativity.
\newblock {\em IEEE Trans. Autom. Contr.}
\newblock In press, doi: 10.1109/TAC.2024.3379452.

\bibitem[Khong and van~der Schaft, 2018]{Khong:18}
Khong, S.~Z. and van~der Schaft, A. (2018).
\newblock On the converse of the passivity and small-gain theorems for
  input-output maps.
\newblock {\em Automatica}, 97:58--63.

\bibitem[Lanzon and Bhowmick, 2023]{Lanzon23}
Lanzon, A. and Bhowmick, P. (2023).
\newblock Characterization of input-output negative imaginary systems in a
  dissipative framework.
\newblock {\em IEEE Trans. Autom. Contr.}, 68(2):959--974.

\bibitem[Liu et~al., 2023]{Liu:23}
Liu, T., Hill, D.~J., and Zhao, J. (2023).
\newblock Incremental dissipativity based synchronization of interconnected
  systems with {MIMO} nonlinear operators.
\newblock {\em IFAC-PapersOnLine}, 56(2):1853--1858.

\bibitem[Megretski et~al., 2011]{MJKR11}
Megretski, A., J\"{o}nsson, U., Kao, C.-Y., and Rantzer, A. (2011).
\newblock The {C}ontrol {H}andbook.
\newblock chapter Integral quadratic constraints. Springer-Verlag, second
  edition.

\bibitem[Megretski and Rantzer, 1997]{MegRan97}
Megretski, A. and Rantzer, A. (1997).
\newblock System analysis via integral quadratic constraints.
\newblock {\em IEEE Trans. Autom. Contr.}, 42(6):819--830.

\bibitem[Moylan and Hill, 1978]{MoyHil78}
Moylan, P.~J. and Hill, D.~J. (1978).
\newblock Stability criteria for large-scale systems.
\newblock {\em IEEE Trans. Autom. Contr.}, 23(2):143--149.

\bibitem[Patra and Lanzon, 2011]{Patra11}
Patra, S. and Lanzon, A. (2011).
\newblock Stability analysis of interconnected systems with ``mixed''
  negative-imaginary and small-gain properties.
\newblock {\em IEEE Trans. Autom. Contr.}, 56(6):1395--1400.

\bibitem[Rantzer, 1996]{Ran96}
Rantzer, A. (1996).
\newblock On the {Kalman-Yakubovich-Popov} lemma.
\newblock {\em Syst. Control Lett.}, 28(1):7--10.

\bibitem[Rantzer and Megretski, 1997]{RanMeg97}
Rantzer, A. and Megretski, A. (1997).
\newblock System analysis via integral quadratic constraints: {Part II}.
\newblock Technical Report TFRT-7559, Lund Institute of Technology.

\bibitem[Ringh et~al., 2022]{RMCQK22}
Ringh, A., Mao, X., Chen, W., Qiu, L., and Khong, S.~Z. (2022).
\newblock Gain and phase type multipliers for structured feedback robustness.
\newblock {\em arXiv preprint arXiv:2203.11837}.

\bibitem[Sassano and Astolfi, 2013]{Sassano:13}
Sassano, M. and Astolfi, A. (2013).
\newblock Dynamic {L}yapunov functions.
\newblock {\em Automatica}, 49(4):1058--1067.

\bibitem[Scherer, 2022]{Sch21}
Scherer, C.~W. (2022).
\newblock Dissipativity and integral quadratic constraints: {T}ailored
  computational robustness tests for complex interconnections.
\newblock {\em IEEE Control Systems Magazine}, 42(3):115--139.

\bibitem[Scherer and Veenman, 2018]{SchVee18}
Scherer, C.~W. and Veenman, J. (2018).
\newblock Stability analysis by dynamic dissipation inequalities: {On} merging
  frequency-domain techniques with time-domain conditions.
\newblock {\em Syst. Control Lett.}, 121:7--15.

\bibitem[Seiler, 2014]{Sei15}
Seiler, P. (2014).
\newblock Stability analysis with dissipation inequalities and integral
  quadratic constraints.
\newblock {\em IEEE Trans. Autom. Contr.}, 60(6):1704--1709.

\bibitem[Sepulchre, 2022a]{Dis22a}
Sepulchre, R. (2022a).
\newblock 50 years of dissipativity theory, {P}art {I}.
\newblock {\em IEEE Control Systems Magazine}, 42(2):6--9.

\bibitem[Sepulchre, 2022b]{Dis22b}
Sepulchre, R. (2022b).
\newblock 50 years of dissipativity theory, {P}art {II}.
\newblock {\em IEEE Control Systems Magazine}, 42(3):5--7.

\bibitem[Sepulchre et~al., 2022]{Sepulchre:22}
Sepulchre, R., Chaffey, T., and Forni, F. (2022).
\newblock On the incremental form of dissipativity.
\newblock {\em IFAC-PapersOnLine}, 55(30):290--294.

\bibitem[Stan and Sepulchre, 2007]{StaSep07}
Stan, G.-B. and Sepulchre, R. (2007).
\newblock Analysis of interconnected oscillators by dissipativity theory.
\newblock {\em IEEE Trans. Autom. Contr.}, 52(2):256--270.

\bibitem[Teel, 1996]{Tee96}
Teel, A.~R. (1996).
\newblock On graphs, conic relations,and input-output stability of nonlinear
  feedback systems.
\newblock {\em IEEE Trans. Autom. Contr.}, 41(5):702--709.

\bibitem[Teel et~al., 2011]{TGPS11}
Teel, A.~R., Georgiou, T.~T., Praly, L., and Sontag, E.~D. (2011).
\newblock Input-output stability.
\newblock In Levine, W.~S., editor, {\em The Control Handbook}, chapter~44.
  Springer-Verlag.

\bibitem[van~der Schaft, 2017]{Sch17}
van~der Schaft, A. (2017).
\newblock {\em $L_2$-Gain and Passivity Techniques in Nonlinear Control}.
\newblock Springer, 3rd ($1$st edition 1996, $2$nd edition 2000) edition.

\bibitem[van~der Schaft, 2021]{Schaft21}
van~der Schaft, A. (2021).
\newblock Cyclo-dissipativity revisited.
\newblock {\em IEEE Trans. Autom. Contr.}, 66(6):2920--2924.

\bibitem[van~der Schaft, 2013]{Schaft13}
van~der Schaft, A.~J. (2013).
\newblock On differential passivity.
\newblock {\em IFAC Proceedings Volumes}, 46(23):21--25.

\bibitem[Verhoek et~al., 2023]{Verhoek:23}
Verhoek, C., Koelewijn, P.~J., Haesaert, S., and T{\'o}th, R. (2023).
\newblock Convex incremental dissipativity analysis of nonlinear systems.
\newblock {\em Automatica}, 150:110859.

\bibitem[Vidyasagar, 2002]{Vid02}
Vidyasagar, M. (2002).
\newblock {\em Nonlinear Systems Analysis}.
\newblock SIAM, 2nd edition.

\bibitem[Willems, 1971]{Wil71}
Willems, J.~C. (1971).
\newblock {\em The Analysis of Feedback Systems}.
\newblock MIT Press, Cambridge, Massachusetts.

\bibitem[Willems, 1972a]{Wil72a}
Willems, J.~C. (1972a).
\newblock Dissipative dynamical systems part {I}: General theory.
\newblock {\em Arch. Rational Mechanics Analysis}, 45(5):321--351.

\bibitem[Willems, 1972b]{Wil72b}
Willems, J.~C. (1972b).
\newblock Dissipative dynamical systems part {II}: Linear systems with
  quadratic supply rates.
\newblock {\em Arch. Rational Mechanics Analysis}, 45(5):352--393.

\bibitem[Willems, 1973]{Wil73}
Willems, J.~C. (1973).
\newblock Qualitative behavior of interconnected systems.
\newblock {\em Annals of Systems Research}, 3:61--80.

\bibitem[Willems, 2007a]{Wil07}
Willems, J.~C. (2007a).
\newblock Dissipative dynamical systems.
\newblock {\em European Journal of Control}, 13(2--3):134--151.

\bibitem[Willems, 2007b]{Wil07history}
Willems, J.~C. (2007b).
\newblock In control, almost from the beginning until the day after tomorrow.
\newblock {\em European Journal of Control}, 13:71--81.

\bibitem[Willems and Trentelman, 1998]{WilTre98}
Willems, J.~C. and Trentelman, H.~L. (1998).
\newblock On quadratic differential forms.
\newblock {\em SIAM J. Control Optim.}, 36(5):1703--1749.

\bibitem[Willems and Trentelman, 2002]{WilTre02}
Willems, J.~C. and Trentelman, H.~L. (2002).
\newblock Synthesis of dissipative systems using quadratic differential forms:
  {Part I}.
\newblock {\em IEEE Trans. Autom. Contr.}, 47(1):53--69.

\bibitem[Zames, 1966]{Zam66}
Zames, G. (1966).
\newblock On the input-output stability of nonlinear time-varying feedback
  systems -- part {I}: Conditions derived using concepts of loop gain,
  conicity, and positivity; and part {II}: Conditions involving circles in the
  frequency plane and sector nonlinearities.
\newblock {\em IEEE Trans. Autom. Contr.}, 11:228--238; 465--476.

\bibitem[Zames and Falb, 1968]{ZamFal68}
Zames, G. and Falb, P.~L. (1968).
\newblock Stability conditions for system with monotone and slope-restricted
  nonlinearites.
\newblock {\em SIAM Journal of Control}, 6(1):89--108.

\bibitem[Zhao et~al., 2022]{Zhao:22}
Zhao, D., Chen, C., and Khong, S.~Z. (2022).
\newblock A frequency-domain approach to nonlinear negative imaginary systems
  analysis.
\newblock {\em Automatica}, 146:110604.

\bibitem[Zhou et~al., 1996]{ZDG96}
Zhou, K., Doyle, J.~C., and Glover, K. (1996).
\newblock {\em Robust and Optimal Control}.
\newblock Prentice-Hall, Upper Saddle River, NJ.

\end{thebibliography}

\end{document}